\documentclass[12pt,a4paper]{article}

\usepackage[T1]{fontenc}

\usepackage[utf8]{inputenc}

\usepackage{a4wide}
\usepackage{latexsym}
\usepackage{epsf}
\usepackage{amssymb}
\usepackage{graphicx}
\usepackage{amsmath, cite}
\usepackage{amsmath,amssymb,amsthm, slashed}
\usepackage{verbatim}
\usepackage{hyperref}
\usepackage{color}	
\usepackage{bbold}

\renewcommand{\d}{\textrm{d}}

\newcommand{\w}{\wedge}

\usepackage{tikz}
\usetikzlibrary{shapes.geometric, arrows}

\newcommand{\x}{\times}

\def\t{\tau}

\newcommand{\be}{\begin{equation}}
\newcommand{\ee}{\end{equation}}
\newcommand{\beq}{\begin{equation}}
\newcommand{\eeq}{\end{equation}}
\newcommand{\ba}{\begin{eqnarray}}
\newcommand{\ea}{\end{eqnarray}}

\renewcommand{\d}{\textrm{d}}

\begin{document}
\numberwithin{equation}{section}

\begin{center}

{\Large {\bf No-scale and scale-separated flux vacua  \\ \vspace{0.4cm}
		  from IIA on G2 orientifolds } }

\vspace{1.1 cm} {\large   Fotis Farakos$^a$, George Tringas$^b$, Thomas Van Riet$^a$}\\

\vspace{0.5 cm} { $a$ Instituut voor Theoretische Fysica, K.U. Leuven,\\
Celestijnenlaan 200D B-3001 Leuven, Belgium.
\\[0.2cm]
$b$ Physics Division, National Technical University of Athens,\\
15780 Zografou Campus, Athens, Greece \\}

\vspace{1cm}
{\bf Abstract}
\end{center}

\begin{quotation}
We discuss flux compactifications of IIA string theory on G2 holonomy spaces with O2/O6-planes to three dimensions and find two classes of solutions: 1) No-scale Minkowski vacua from NSNS 3-form fluxes and RR 4-form fluxes. 
2) By adding Romans mass we find AdS$_3$ vacua for which the AdS scale can be decoupled completely from the KK scale while the solution is at tunable weak coupling and large volume. For the AdS$_3$ vacuum we only have a proper 3D description (i.e. smeared orientifold description) of the solution like the 4D analogue of IIA moduli stabilization from fluxes and O6-planes. This 3D description reveals that moduli with non-compact moduli spaces can be stabilized at the classical level. For both types of vacua we can have supersymmetry or not. 
\end{quotation}
\newpage
\tableofcontents 
\newpage

\section{Introduction}

Moduli stabilization for string compactifications \cite{Ibanez:2012zz,Silverstein:2004id} is one of the most relevant activities in string phenomenology. For example, without stabilized moduli one does not reproduce 4-dimensional relativity since massless (or very light) scalar fields mediate forces in such a way that the equivalence principle is broken. A crucial ingredient for stabilizing moduli in a computable way are fluxes together with orientifold planes and we refer to \cite{Ibanez:2012zz,Silverstein:2004id,Grana:2005jc, Douglas:2006es, Denef:2007pq} for reviews on that topic. It is fair to say that by now, our best understanding of flux compactifications involve compactifications to AdS space, whereas compactifications to dS space are notoriously difficult to control, if they exist at all \cite{Danielsson:2018ztv, Cicoli:2018kdo}. Compactifications down to Minkowski space with stabilized moduli are not known to us.

Another very basic feature of string compactifications with stabilized moduli is that the compact dimensions should be small enough in order to find a genuine 4-dimensional effective field theory. Usually one defines ``small'' with respect to the 4-dimensional Hubble scale. This feature is also referred to as scale separation since the KK scale is supposed to be parametrically separated from the Hubble scale. 

In this paper we focus on AdS compactifications. So far our best motivated scenarios for achieving moduli stabilization in scale-separated AdS vacua are \cite{Kachru:2003aw, Balasubramanian:2005zx, DeWolfe:2005uu}, where only reference \cite{DeWolfe:2005uu} succeeds in finding an arbitrary separation between the KK scale and the AdS scale by taking an unconstrained flux quantum to infinity. On the downside this AdS vacuum does have scalars of the order of the AdS mass scale and hence these vacua are not so useful for true phenomenological models. 

Interestingly, the very basic assumption of being able to achieve scale separation is under debate \cite{Gautason:2015tig, Lust:2019zwm, Gautason:2018gln, Blumenhagen:2019vgj, Font:2019uva} since the compactifications used in string phenomenology always feature ingredients which obscure fully explicit computations. One can hope that holography could settle this discussion once it can be shown there are huge families of CFTs that could reproduce qualitative features like full moduli stabilization and scale separation. We refer to \cite{Polchinski:2009ch, deAlwis:2014wia, Conlon:2018vov, Alday:2019qrf, Perlm, Buratti:2020kda} for recent work in that direction.

These issues motivate us to investigate moduli stabilization and scale separation in 3D vacua. Such vacua or not obviously relevant for phenomenology but the ingredients used in the construction of 4D vacua exist there as well, whereas that seems not to be true in 5D or higher in our opinion. This already makes 4 dimensions special in string theory. The nice feature of 3 dimensions is that supersymmetric AdS vacua can be dual to SCFTs in 2D. Since 2D CFTs are somewhat more studied than 3D CFTs our hope is that we can settle the issue for 3D AdS vacua\footnote{One can of course also ponder AdS vacua in 2D, see for instance \cite{Lust:2020npd}. }. Although, in this paper we focus entirely on the supergravity problem and do not venture into a holographic description. We hope to come back to this in the future. 

So we are led to consider compactifications of string theory on seven-dimensional manifolds that preserve some amount of supersymmetry. 
Furthermore, we want to have some handle on the moduli problem for such manifolds.  This restricts us to manifolds with G2 holonomy, 
since by now there is an extended literature on the moduli problem for such spaces; 
see e.g. \cite{Beasley:2002db, Acharya_2002, Gutowski:2001fm, Karigiannis_2005, GRIGORIAN_2010} for a sample of papers. 
In contrast, 
manifolds with a G2 structure group but no G2 holonomy are much less understood when it comes to fluctuation theory. 

Spaces with G2 holonomy, abbreviated G2 spaces from here onwards, are Ricci flat and compactifying over them without fluxes and sources will lead to a 3D supergravity theory with 4 supercharges and a moduli space. 
We are interested in lifting the moduli space using fluxes 
and therefore one should worry whether the flux backreaction will drive the system away from the Ricci flat internal G2 space. 
It is believed that orientifolds ameliorate this issue to some extend as we explain below. 
Orientifolds are anyhow needed to cancel the RR tadpoles induced by the fluxes and also to circumvent certain nogo arguments for having Minkowski solutions or scale-separated AdS solutions\footnote{Scale separation here means a decoupling of the KK scale from the AdS scale \cite{Tsimpis:2012tu}.} \cite{Gautason:2015tig}.

The negative tension of orientifold planes can help in keeping the deformation away from the Ricci flat manifold (G2, Calabi--Yau) under control. This is best seen when the orientifolds are smeared over the internal dimensions as to reflect the course graining over distances smaller than the KK scale. 
Smeared orientifolds provide the necessary negative energy/momentum in order to cancel the positive energy/momentum of the fluxes, 
thus providing a well behaved ``solution'' with a flat internal space \cite{Grana:2006kf, Acharya:2006ne, Blaback:2010sj}. 

Of course orientifolds are localized objects in string theory and if one wants a more sensible solution that can be probed at distances smaller than the KK scale one needs to find 10D supergravity solutions with orientifold singularities. 
This can be done explicitly for flux solutions that do not involve intersecting planes, but only parallel ones \cite{Grana:2006kf, Blaback:2010sj, Andriot:2016ufg}. 
Such solutions have internal manifolds that differ from the smeared solution, and furthermore have a warp factor in front of the external space metric. Nonetheless the backreaction of the orientifolds is mild since the alteration of the internal metric can be described in terms of a conformal factor multiplying the space transversal to the wrapped planes\footnote{To be more precise: parallel branes and planes wrap a fibre of a bundle. 
The conformal factor of the base equals the warp factor and the conformal factor of the fibre equals the inverse in string frame. }. 
From the 10D equations one can then verify that derivatives of the warp and conformal factor are crucial in canceling the flux energy/momentum. 
Nonetheless the smeared solution is well approximated away from the sources, especially in the limit of weak coupling and large volume \cite{Baines:2020dmu}. This can be understood from the fact that those limits dilute the fluxes sufficiently such that the original flat space Ansatz was sensible. This connection between smeared and localized solutions is not proven for flux vacua with intersecting sources but there is some recent evidence that points towards it \cite{Junghans:2020acz, Marchesano:2020qvg}.

Crucially, the low energy effective field theories that are commonly used for flux compactifications do not take into account warping and other backreaction effects, so they effectively probe the smeared orientifold solutions. This is not strange and just means that the EFT course grains over distances smaller than the KK scale. 
There are constructions however which do take it into account and go under the name of \emph{warped effective field theory}; 
see \cite{Giddings:2005ff, Douglas:2008jx, Shiu:2008ry, Martucci:2014ska, Koerber:2007xk, Frey:2013bha, Frey:2008xw} for a biased sample of papers on the topic. 

Four-dimensional flux vacua of interest to phenomenology that are obtained within the classical realm of 10D supergravity 
with D-brane and O-plane sources come in two kinds: 
1) there are the no-scale Minkowski vacua in IIB from 3-form fluxes and their T-duals \cite{Giddings:2001yu, Dasgupta:1999ss, Grana:2006kf} 
or 2) the scale-separated AdS vacua in IIA \cite{DeWolfe:2005uu, Derendinger:2004jn} and their IIB cousins \cite{Caviezel:2009tu, Petrini:2013ika}. 
Whereas the no-scale Minkowski solutions tend to feature parallel sources and their 10D description in terms of localized orientifolds is more or less understood, 
the same does not hold for the scale-separated AdS vacua, which always feature intersecting sources. 
For the latter we are also not aware of an EFT improvement, like warped effective field theory, to describe the backreaction effects in the scale-separated AdS vacua. 
Although as we mentioned earlier there is recent progress towards justifying the smeared approach \cite{Junghans:2020acz, Marchesano:2020qvg}. 

In this paper we follow the ``standard'' strategies used in compactifications to 4D:  we are led to find vacua in 3D using fluxes and sources as described directly in the 3D supergravity. 
Similarly to 4D, we find the two classes: no-scale Minkowski vacua and scale-separated AdS vacua, and again it is only clear for the no-scale vacua how to ``backreact'' the orientifolds properly and find 10D solutions.

\section{Fluxes and sources on G2 spaces}\label{sec:possible}

Before we dive into concrete flux compactifications we wish to understand what the possibilities are, 
insisting that the internal manifold is of G2 holonomy (in the smeared limit). Note that the expected 3D supergravity theories will only have 2 real supercharges. The reason is that G2 holonomy already restricts to 4 real supercharges and the orientifolds further cut them in half. 

To list the possibilities we work in the democratic formalism for the fluxes and then take all fluxes to be internal (also called magnetic), 
without loss of generality. For instance an electric $H_3$ flux is then simply described by magnetic $H_7$ flux.  All fluxes we can have are
\begin{align}
& \text{IIA}\,:\, H_3, H_7, F_0, F_2, F_4\,,\nonumber\\
& \text{IIB}\,:\, H_3, H_7, F_3, F_5, F_7\,,\nonumber
\end{align}
where we used that a G2 space has no one- or six-cycles and we assume that fluxes are closed but non-exact. 

For reasons explained in the introduction, 
we make sure that the O-plane charges in the compactification are always cancelled by fluxes and not only by D-branes. 
So we will ignore D-branes in what follows, although they will eventually be impossible to avoid in some of our concrete models. 
The relevant Bianchi identities and flux EOM are summarized by
\be
\d F_{8-p} = H_3 \wedge F_{6-p} + \delta_{{\rm O}p}\,.
\ee
The possible planes that can be used, bearing in mind there are no one- or six-cycles, are
\begin{align}
& \text{IIA}\,:\, \text{O}2, \text{O}4, \text{O}6\,,\nonumber\\
& \text{IIB}\,:\, \text{O}5, \text{O}7\,.\nonumber
\end{align}
The charges of O7 planes or O5 planes can never be cancelled by fluxes. For O7 planes this is obvious and for O5 planes we have to realize that the Bianchi identity requires non-zero $F_1$ flux, which cannot be present because of a lack of one-cycles.  This rules out IIB altogether.  We are now left with IIA. 

We ignore O4 planes all together since they wrap 2-cycles and G2 has calibrated 2-cycles, but they are never supersymmetric \cite{Gutowski:2001fm}. 
We thus have O2 and O6 left.

Consider an O2 filling the 3D space. The parities of the fluxes are
\begin{align}
& \text{even}\,:\, F_0, F_4, H_7\,,\nonumber\\
& \text{odd}\,:\, F_2, H_3\,.\nonumber
\end{align}
But there are no odd two-forms in the space transversal to the O2, neither even seven-forms, so $H_7$ and $F_2$ are removed. 
Then, \emph{in absence of O6 planes} we also have to remove $F_0$ because of the $F_2$ Bianchi identity (note that we do not want to eliminate $H_3$ since then we cannot cancel the O2 tadpole), 
which brings us to 
\be
\text{O2 allowed flux:}\quad F_4, H_3\,.\nonumber
\ee
The solutions in this model have been considered earlier in \cite{Blaback:2010sj} and are of the no-scale Minkowski type where $\star_7 H_3\sim F_4$. If the internal space would be $S^1\times CY_6$ this would be T-dual to the well-known 4D Minkowski solutions in IIB with 3-form fluxes. The ``solutions'' described in \cite{Blaback:2010sj} are completely general and only the conditions to solve the 10D equations were stated. Neither supersymmetry nor the moduli problem was treated. One of the aims of this paper is to fill this gap and construct the 3D supergravity in case the internal manifold is G2.  

Since space-filling O2-planes can intersect space-filling O6-planes in a SUSY manner, we can consider their combination. 
Then the above reasoning goes through but without removing the $F_0$ flux. 
So we have 
\be
\text{O2/O6 allowed flux:}\quad F_0, F_4, H_3\,.\nonumber
\ee

Imagine we want to have an O6-plane and cancel its tadpoles with fluxes and that we do not want an O2-plane. 
O6-planes wrap 4-cycles inside the G2 space that need to be calibrated in a supersymmetric manner if one wants to achieve a 3D supergravity description.  Note that $H_3\wedge F_2$ and $H_3\wedge F_4$ then have to vanish from the Bianchi identities. 
From the tadpole cancellations we get $F_2 = 0$, whereas we can keep $F_4 \ne 0$ as long as it wedges to zero with the $H_3$, 
which brings us to 
\be
\nonumber 
\text{O6 allowed flux:}\quad F_0, F_4, H_3 \,, \quad F_4 \w H_3 \equiv 0  \, . 
\ee

To summarize, 
assuming G2 holonomy we could prove that the only models in the market are O2, O6 and possibly O2/O6 together.

\section{The no-scale model}

In this section we discuss the ``no-scale'' type backgrounds obtained from space-filling O2-planes whose tadpole is cancelled by a combination of $F_4$ and $H_3$ fluxes that obey the proportionality rule $H_3\sim \star F_4$. The 10D description of such backgrounds has appeared in \cite{Blaback:2010sj}, 
but without any details; 
rather, it was only explained what is needed to solve the 10D equations of motion. 
Here we discuss how to obtain the effective field theory in 3D when warping can be sufficiently ignored. 
These backgrounds are inspired from T-duality of the GKP backgrounds in IIB \cite{Giddings:2001yu}.

Because we compactify on G2 spaces we expect maximally four of the original 32 supercharges to be preserved. 
The O2-plane further reduces the number of supersymmetries in half (i.e. two supercharges remain) by essentially truncating the spectrum. 
As a result the low energy theory will be a 3D N=1 supergravity. 
Our interest here is in finding the exact low energy description of such theory by matching the 3D objects to the 10D data. We first start with a general discussion of 3D minimal supergravity.

\subsection{Brief summary of minimal 3D supergravity} \label{sec:3Dsugra}
Here we list few basic aspects of three-dimensional N=1 supergravity and later we will present the specific theory of our interest. 
More details can be found in Appendix \ref{app:3D}. 

The bosonic sector of N=1 supergravity in 3D has a metric field, real scalar fields $\phi^I$ and (abelian) vectors $A^{(A)}$, which can be dualized to scalars. 
The bosonic part of the general action is 
\be
\label{sugra3D}
e^{-1} {\cal L} = \frac12 R 
- g^{\mu \nu} G_{IJ}(\phi) \partial_\mu \phi^I \partial_\nu \phi^J 
- \frac14 f(\phi) F_{\mu \nu}^{(A)} F^{\mu \nu (A)} - V(\phi) \, ,  
\ee
with 
\be\label{Pfunc}
V(\phi) = G^{IJ}  P_I P_J  - 4 P^2 \, . 
\ee 
We name the function $P$ the real superpotential and $P_I$ is shorthand for $\partial_I P$. 
The gauge kinetic function $f(\phi)$ is real but otherwise unrestricted. 
From here onwards $G^{IJ}$ is the inverse of the target space metric $G_{IJ}$.  

On the fermion side we have the gravitino $\psi_{\mu}$, the Majorana spinors $\chi^I$ which are the superpartners of the real scalars $\phi^I$, 
and the superpartners of the vectors (gaugini) which are also Majorana and denoted $\lambda^{(A)}$. 
We need the supersymmetry variations in order to understand the supersymmetry invariance of flux solutions 
\be
\begin{aligned} \label{SUSYvari}
	\delta \psi_\mu \Big{|}_{\rm shift} & = \frac14 \omega_\mu^{ab} \gamma_{ab} \epsilon 
	- P \gamma_\mu \epsilon \, , 
	\\
	\delta \chi^I \Big{|}_{\rm shift} & = G^{IJ} P_J \epsilon \, , 
	\\
	\delta \lambda^{(A)} \Big{|}_{\rm shift} & = 0 \, , 
\end{aligned}
\ee
where $\epsilon$ is the 2-component fermionic Majorana local supersymmetry parameter.

\subsection{A 10D view on the effective theory} 

We will work with IIA supergravity with space-filling O2-planes and we will have only non-vanishing $F_4$ and $H_3$ fluxes as dictated by
\be
\label{tadO2}
\d F_6 = H_3\wedge F_4 + Q\delta_7 \,, 
\ee  
where $Q$ is the O2 charge. 
The relevant bosonic part of the 10D action is (in 10D Einstein frame) 
\begin{equation}
\label{10Dstart}
S = \int_{10} \left(\star_{10}R - \tfrac{1}{2}e^{-\phi}\star H_3 \wedge H_3 -\tfrac{1}{2}e^{\phi/2}\star F_4\wedge F_4\right) 
+ e^{-\phi/4}\mu\int_{{\rm O}2}\sqrt{|g_3|} \, . 
\end{equation}
The O2 plane has its world-volume perpendicular to the 7-dimensional internal G2 space on which we compactify on. 

From the 10D action one can readily find the 10D form of the scalar potential in 3D as we now show. 
A direct dimensional reduction gives the following 10D expression for the 3D potential $V$ given by 
\begin{equation}
V = \int_7 \left( \tfrac{1}{2}e^{-\phi}\star H_3 \wedge H_3 +\tfrac{1}{2}e^{\phi/2}\star F_4\wedge F_4 - e^{-\phi/4}\frac{\mu}{Vol_7}\epsilon_7   \right)\,, 
\end{equation}
where now the stars are 7D Hodge stars and $\epsilon_7$ is the 7D volume form. Because the O2-planes are BPS objects their tension and charge are related as  $\mu=\pm Q$  where the minus sign refers to anti-O2-planes. 
That charge, however, can be obtained from the Bianchi identity \eqref{tadO2}, or equivalently from the RR tadpole condition 
\be
\label{tadtad}
-\int H_3 \wedge F_4 = Q = \int_7 \frac{Q}{Vol_7}\epsilon_7\,.
\ee
As a result the effective potential becomes 
\begin{equation}
V = \int_7 \left( \tfrac{1}{2}e^{-\phi}\star H_3 \wedge H_3 +\tfrac{1}{2}e^{\phi/2}\star F_4\wedge F_4 \mp e^{-\phi/4} H_3\wedge F_4 \right)\,, 
\end{equation}
which can be recast in a form that is a manifest total square\footnote{In our notation the square of a form is defined as: 
$\sqrt{g}F^2 d^7x = 
\star F\wedge F = 
\sqrt{g}\frac{1}{p!}F_{\mu_1\ldots \mu_p}F^{\mu_1\ldots \mu_p}  d^7x $} 
\begin{equation}
\label{scasca}
V = \frac{1}{2}\int_7 \sqrt{g_7}\left(e^{-\phi/2} H_3\mp e^{\phi/4}\star_7 F_4 \right)^2\,.
\end{equation}
One of the $\pm$ signs corresponds to O2-planes while the other to anti-O2-planes. 
Since the scalar potential is a total square, 
the vacuum must live at configurations for which the square vanishes, 
that is 
\be
\label{vacvac}
H_3\mp e^{3\phi/4}\star_7 F_4=0 \, , 
\ee
and therefore the solutions are Minkowski. 
Whether supersymmetry is present or not is discussed below.

Let us now discuss the spectrum of bosonic fields in 3D. The graviton of course is given directly by the dimensional reduction of the 10D metric tensor,  so let us now focus on what can be the scalars of the 3D theory.  Recall that in going from 11D to 4D on a G2, 
see e.g. \cite{Beasley:2002db},  the scalars come from the metric and the gauge three-form $C$ of the 11D supergravity, 
which are decomposed as 
\be
C = c^i \Phi_i + \dots \ , \quad \Phi = s^i \Phi_i \, , 
\ee 
where the $s^i$ and the $c^i$ are real scalar moduli and where $\Phi$ is the closed and co-closed three-form invariant under G2 transformations. 
The $\Phi_i$ span a harmonic basis of $b_3$ three-forms with $b_3$ the 3rd Betti number of the G2 space:
\be
\label{3basisGEN}
\{ \Phi_i \} \ , \quad  i = 1 , \dots , b_3 \ , \quad \Phi = s^i \Phi_i \,. 
\ee
The real scalars combine into complex holomorphic moduli as: $z^i = c^i + i s^i$. 
The low-energy theory is standard 4D N=1 supergravity and the kinetic terms of this sector are 
\be
{\cal L}_{kin.} = - g_{i\bar j} \partial_\mu z^i \partial^\mu \overline z^{\bar j} \, , 
\ee
where the K\"ahler metric is given by 
\begin{equation}
g_{i\bar j} = \frac14 \text{vol}(X)^{-1} \int_X \Phi_i \wedge\star \Phi_j \, . 
\end{equation}  
Let us now turn to compactifications of IIA to three dimensions. 
At first sight we would expect to have moduli from $C_3$, $C_1$ and $B_2$. 
However, since there are no one-cycles, there are no scalars coming from reducing $C_1$. 
There is also no scalar coming from dualizing the vector $C_1$ to a scalar in 3D since the vector is projected out by the O2-plane, 
thus the dilaton also remains a genuine real scalar. 
Indeed, without the orientifold, that dualized vector would have paired up with the dilaton to a complex field. 
Moreover there are no scalars coming from $C_3$ either, again because of the O2 projection since all 3-cycles are necessarily odd whereas $C_3$ is even under O2.  There are furthermore no scalars from reducing $B_2$ over two-cycles because there are no odd 2-cycles for an O2 here.  
There are however even two-cycles and reducing $C_3$ over them gives vectors in 3D that dualize to scalars. 
These are actually axions with compact field ranges. 
Without the O2 projections they would have paired up with the scalars from $B_2$ to give complex fields.  
We therefore find that exactly those scalars that would have paired up with the metric scalars ($s^i$) to give complex scalars are absent consistent with the real formulation of minimal 3D supergravity.  
 
To sum it up, all scalars come from the metric (ignoring for the moment the dilaton) which one obtains from expanding $\Phi$ over a basis of harmonic 3-forms. So we have $b_3$ amount of metric scalars $s^i$ and $b_2$ axions from $C_3$ expanded along harmonic two-forms. 
There can also be scalars corresponding to D2 or D6 positions if D2- or D6-branes are needed for tadpole cancellation. 
Also such scalars will have compact field ranges. 

Our fluxes $F_4$ and $H_3$ can give masses to the metric scalars but not to the axions and brane positions, which instead should get a mass from quantum effects. 
However, we will find that a linear combination of the volume and the dilaton necessarily remains massless in the Minkowski vacua where only \eqref{vacvac} holds.  

Indeed, at this point we can readily study the generic formulas that arise from our discussion,  and ask how many moduli can be stabilized from imposing \eqref{vacvac}.  In the spirit of this section we use generic G2 ingredients instead of restricting to specific models. We denote a generic G2 space as $X$ and introduce the basis for the dual 4-form $\Psi=\star \Phi$ on $X$ as 
\be
\label{4basisGEN}
\{ \Psi_i \} \ , \quad  i = 1 , \dots , b_3 \ , \quad \int \Phi_i \wedge \Psi_j = \delta_{ij} \, . 
\ee
We also define
\be\label{GIJ}
G_{ij} = \frac{7}{4} \frac{\int \Phi_i \wedge \star \Phi_j}{\int \Phi \wedge \star \Phi} = 
\frac14 \text{vol}(X)^{-1} \int \Phi_i \wedge \star \Phi_j   \, , 
\ee
where we used $\text{vol}(X) = \frac17 \int \Phi \wedge \star \Phi$. The bilinear form $G_{ij}$ becomes a metric on the moduli space. We will expand on that in the next subsection. Then the expansion of the $\star \Phi_i$ in the 4-form basis \eqref{4basisGEN} generically can be expressed as 
\be
\label{BIJ}
\star \Phi_i = B_i{}^{l}(s) \, \Psi_l \, , 
\ee
and by taking the wedge product with $\Phi_j$ and using the orthonormality in \eqref{4basisGEN} we find 
\be
\label{BIJG}
B_i{}^l(s) = 4 \text{vol}(X) G_{ij} \delta^{jl} \, , 
\ee
which means that the moduli-dependent coefficients $B_i{}^l(s)$ are completely specified by the geometry. 
Note that $B_i{}^l(s)$ is a $b_3(X) \times b_3(X)$ invertible matrix. 

We expand the fluxes as
\be 
F_4= f^i \Psi_i \ , \quad H_3 = h^i \Phi_i \, , 
\ee 
with $f^i$ and $h^i$ flux quanta that should be properly quantized. 
If we insert them into  \eqref{vacvac} together with \eqref{BIJ} and \eqref{BIJG} we find 
\be
h^i B_i{}^k \Psi_k = \pm e^{3 \phi/4} f^k \Psi_k \ \rightarrow \   h^i B_i{}^{j}(s) = \pm e^{3 \phi/4} f^j \, . 
\ee
We see that we have in principle $b_3$ conditions that can thus fix up to $b_3$ moduli, 
which are essentially all the $s^i$. 
We will show below that a linear combination of dilaton and volume  remains free and has to be fixed by different mechanisms.  The flux parameters $f^i$ and $h^i$ are a total of $2 b_3 -1$ independent parameters  because of the tadpole condition \eqref{tadtad} which requires 
\be
f^i \delta_{ij} h^j = - Q \, . 
\ee
One can therefore expect that all of the $s^i$ will be fixed in terms of the dilaton because there is a large freedom in choosing the fluxes.  Of course for special values of the $f^i$ and $h^i$ not all $s^i$ will be fixed.

\subsection{The 3D supergravity effective theory} 

Let us now turn to the specific low-energy supergravity theory we want to study.  
For compactifications of IIA on Calabi--Yau orientifolds with fluxes the 4D N=1 supergravity theory was constructed in \cite{Grimm:2004ua}, 
here we perform a similar investigation for IIA on G2 orientifolds. 
Since we readily have the scalar potential given by \eqref{scasca}, we need also the kinetic terms in order to match the 3D supergravity with the compactified effective theory. 
In 10D Einstein frame,  the reduction Ansatz for the metric is
\be
\label{10DmetricV}
ds^2_{10} =  e^{2\alpha v} ds^2_3 + e^{2\beta v}\widetilde{ds}^2_7\,,
\ee
where $v$ is a 3D scalar that accounts for the volume and hence $\widetilde{ds}^2_7$ is the metric on a unit-volume G2 space.  With the specific choice of numbers $\alpha^2 = 7/16$ and $-7\beta =\alpha$ we find canonical kinetic terms in 3D 
\be
\label{kinetic+V}
e^{-1}\mathcal{L} = R_3 - \tfrac{1}{2}(\partial v)^2 -  \tfrac{1}{2}(\partial \phi)^2 - V\,. 
\ee
Taking now into account the volume scalings, 
the expression for the potential \eqref{scasca} takes the form 
\begin{equation}
V = \frac{1}{2}\int_7 \sqrt{\tilde{g}_7} e^{-21\beta v}\,\left(e^{-\phi/2} e^{\beta v/2} H_3\mp e^{-\beta v/2}e^{\phi/4}\tilde{\star}_7 F_4 \right)^2 \, , 
\end{equation}
where $\tilde{g}_7$ is the unit-volume metric. 
Finally we consider the following \emph{orthonormal} redefinition of 
scalars\footnote{For later convenience we also note that $\phi = - \frac{3 \sqrt 7}{8} x - \frac18 y$ 
	and $7 \beta v =\frac{\sqrt 7}{32} x - \frac{21}{32} y$.} 
\be
\label{redefs}
\frac{x}{\sqrt7} = - \frac{3\phi}{8} + \frac{\beta}{2}v\,,\qquad 2y = -21\beta v - \frac{1}{4}\phi \,,  
\ee
and the action takes the form 
\be\label{lagr}
e^{-1}\mathcal{L} = R_3 - \tfrac{1}{2}(\partial x)^2 -  \tfrac{1}{2}(\partial y)^2 +\ldots - V\,, 
\ee
where the $\ldots$ denote kinetic terms for all other geometric scalars and 
\begin{equation}
\label{VTVR}
V = \frac{1}{2}\int_7 \sqrt{\tilde{g}_7} e^{2y}\,\left(e^{\frac{x}{\sqrt7}} H_3\mp e^{-\frac{x}{\sqrt7}}\tilde{\star}_7 F_4 \right)^2\,. 
\end{equation}
In the rest of this section we will derive the effective 3D supergravity theory that gives rise to this action, meaning we fill in the dots of equation (\ref{lagr}). Note already that $y$ is the massless ``no-scale'' direction we mentioned earlier and it indeed corresponds to a linear combination of dilaton $\phi$ and volume $v$. 

Let us also establish the normalization of the kinetic terms for the volume-preserving fluctuations,  as it will be helpful to cross-check our results later.  For that we consider a single extra fluctuation, 
say $z$,  of a unit-volume torus that serves as a proxy for our G2 internal space.  We will also keep the volume modulus $v$ in our discussion because we want to keep track of the relative normalizations.  We have for the unit-volume metric 
\begin{equation}
\label{unit-torus}
\widetilde{d s}_7^2 =  e^{2\xi z}[dy_1^2 + dy_2^2 + dy_7^2] +  e^{2\delta z}[dy_4^2 + dy_5^2 + dy_6^2 + dy_3^2]\,, 
\end{equation}
where unit volume implies $3\xi=-4\delta$.  Direct dimensional reduction of the 10D Einstein--Hilbert term gives 
\begin{equation}
\label{kinetic-z} 
\int_3 \sqrt{-g_3}\left(R_3 -\tfrac{1}{2}(\partial v)^2  -\tfrac{21}{4}\xi^2(\partial z)^2 \right)\,, 
\end{equation}
therefore we  set $\xi^2 = 2/21$ to have canonical kinetic terms. 
We will use this normalization as a way to double check later our kinetic terms for the $v$ the $s^i$ moduli. 
We can express the relevant part of the G2 form, keeping only the modulus $z$ while freezing the rest, as 
\be
\label{zPhi1}
\tilde \Phi = e^{3\xi z}dy^{127} + \dots \ , \quad \tilde{\star} \tilde \Phi =  e^{4\delta z}dy^{3456} + \dots \, , 
\ee 
where $dy^{127} = dy^1 \wedge dy^2 \wedge dy^7$ and $dy^{3456}=dy^{3}\w dy^{4}\w dy^{5}\w dy^{6}$. 
If we wish to generate this term from a general formula using the $\Phi$-form we find that it comes from
\be\label{formula}
e^{-1}\mathcal{L}_{\text{kin}} = -\frac{1}{2}\text{vol}(X)^{-1}\int_7 \Phi_i\wedge\star\Phi_j \partial s^i\partial s^j\,.
\ee
This is consistent with the kinetic term one derives in going from 11 to 3 \cite{Beasley:2002db}\footnote{There is a 1/2 factor different from \cite{Beasley:2002db} but can be understood from the different normalization of the Einstein Hilbert term. }. 
But there is a small subtlety compared with the literature on G2 compactifications from 11D. 
For that, let us use that 
\be
s^i = e^{3\beta v} \tilde{s}^i
\ee
with $\tilde{s}^i$ the fluctuations for unit-volume spaces. Equation (\ref{formula}) contributes the following piece, $-\frac{9}{32}(\partial v)^2$, 
to the kinetic term for the volume. Hence we deduce that the full kinetic term should be 
\be
\begin{aligned}
\label{formula2}
e^{-1}\mathcal{L}_{\text{kin}} =&  -\frac{1}{2}(\partial\phi)^2 - \frac{7}{32}(\partial v)^2 -\frac{1}{2}\text{vol}(X)^{-1}\int_7 \Phi_i\wedge\star\Phi_j \partial s^i\partial s^j 
\\ 
=& -\frac{1}{2}(\partial\phi)^2 - \frac{1}{2}(\partial v)^2 -\frac{1}{2}\text{vol}(\tilde X)^{-1}\int_7 \Phi_i\wedge \tilde \star\Phi_j \partial \tilde{s}^i\partial \tilde{s}^j\,.
\end{aligned}
\ee
One can verify that in contrast to our situation, in going from 11 to 4, the contribution of (\ref{formula}) to the volume kinetic term is complete. 
Notice also that in the second line of \eqref{formula2} we have used the tilde notation to refer to the G2 expressions 
that instead of the $s^i$ make use of their unit-volume counter-parts, 
the $\tilde s^i$. 
For the later in fact it holds that $\text{vol}(\tilde X)=1$, but we have chosen to keep the expressions with the unit volume manifest to avoid any confusion.

From here onwards we change to different normalizations to make contact with the literature on 3D gravity. This we can simply do by rescaling the 3D metric as $g_{\mu\nu}\rightarrow \frac{1}{4}g_{\mu\nu}$, such that we end with
\begin{equation}
\label{finalaction} 
e^{-1}\mathcal{L}_{\text{kin}} = \tfrac{1}{2}R_3 -\tfrac{1}{4}(\partial x)^2  -\tfrac{1}{4}(\partial y)^2  -\tfrac{1}{4}\text{vol}(\tilde X)^{-1}\int_7 \Phi_i\wedge\tilde\star\Phi_j \partial \tilde{s}^i\partial \tilde{s}^j \,.  
\end{equation}
In the language of subsection \ref{sec:3Dsugra} we have the real scalars $\phi^I = x,y,\tilde s^i$. The $x$ and $y$ are a combination of the volume modulus and of the dilaton,  whereas the $\tilde{s}^i$ are moduli for the G2 metric deformations (but not volume) and the sigma model metric of (\ref{sugra3D}) is given by 
\be
\label{scalarmetric}
G_{IJ} =   \begin{bmatrix}
    1/4 & 0 & 0 \\
    0 & 1/4 & 0 \\
    0 & 0 & G_{ij}
  \end{bmatrix} \,, 
\ee 
with $G_{ij}$ defined earlier in (\ref{GIJ}) where now all $s^i$ are replaced with $\tilde s^i$. 
But note that this is overcounting the degrees of freedom since the above notation seems to imply that the $\tilde{s}^i$ are independent whereas they should multiply to a fixed number as they describe the unit-volume G2 space - indeed we know that after all $\text{vol}(\tilde X)=1$. 
In what follows we will demonstrate that we can pretend the $\tilde{s}^i$ to be independent and at the very end use that their product equals one. 
This is fully consistent with all our calculations in the bosonic sector and also with supersymmetry as we shown in the appendix.

To fix the 3D supergravity theory we need to find the real superpotential function $P$, which will inform us about the supersymmetry shifts via equation (\ref{SUSYvari}). Instead of doing a full derivation using 10D supersymmetry rules, 
we will guess the answer for $P$ and verify that indeed $P$ leads to the scalar potential (\ref{VTVR}) we derived already from 10D, 
through equation (\ref{Pfunc}).  
The answer for $P$ can be guessed in analogy with existing superpotentials for flux compactifications to be 
\be
\label{SUP}
P =  \frac{e^y}{8} \left[ \gamma e^{x/\sqrt{7}} \int \star\Phi \wedge H_3 \, \text{vol}(X)^{- \frac47} 
+ e^{-x/\sqrt{7}} \int \Phi \wedge F_4 \, \text{vol}(X)^{- \frac37}  \right] \, . 
\ee 
The number $\gamma$ will turn out to be $\pm 1$, depending on whether we look at theories with O2 planes or anti-O2-planes. 
The reader may notice that we have used the $s^i$ in \eqref{SUP} instead of the $\tilde s^i$, 
however, 
it is easy to check that $P(s^i)\equiv P(\tilde s^i)$. 
When we take derivatives of $P$ with respect to $\tilde s^i$, 
i.e. $P_i$, 
we will of course use the $P(\tilde s^i)$ version.

Now we want to evaluate the scalar potential. 
As we explained the 3D supergravity built from $G_{IJ}$ and $P$ will contain one additional degree of freedom because of the double counting of  the volume that will in any case appear in the kinetic terms and from the one additional fermionic field.  For the moment we ask the reader to bear with the extra degree of freedom until we show that it can be fixed consistently.  
So our approach here is to simply verify that our choice of $P$ leads to the correct scalar potential derived from 10D. 

We first note that $y$ is special since
\be
(G_{yy})^{-1} P_y P_y - 4 P^2 = 0 \, , 
\ee
and as a result the scalar potential takes the from 
\be
\label{Vone}
V = G^{ij} P_i P_j + (G_{xx})^{-1} P_x P_x  \, . 
\ee
This is what we call no-scale in 3D. First we evaluate $P_x$ which gives 
\be
\label{hPx}
P_x  = \frac{\partial P}{\partial x} =  
- \frac{1}{\sqrt 7} \frac{e^y}{8} \int \text{vol}(\tilde X)^{- \frac37}  \, \tilde \Phi \wedge {\cal F}_4 \, ,  
\ee
where we remind the reader that we use the notation $\tilde \Phi = \Phi_i \tilde s^i$, etc. and 
\be
\label{calF4}
{\cal F}_4 = e^{- x / \sqrt 7} F_4 - \gamma e^{x / \sqrt 7} \tilde \star H_3 \, \text{vol}(\tilde X)^{- \frac17}  \, . 
\ee 
Therefore we find 
\be\label{Px^2}
(G_{xx})^{-1} P_x P_x  = \frac47 \left( \frac{e^y}{8} \right)^2 \left( \int \text{vol}(\tilde X)^{- \frac37}  \, \tilde \Phi \wedge {\cal F}_4 \right)^2 \, . 
\ee
Now we want to evaluate the $P_i$. 
To do this we need a series of properties that we now list. 
The derivatives with respect to $\tilde{s}^i$ are defined as 
\be
\frac{\partial}{\partial \tilde{s}^i} \Phi = \Phi_i \ , \quad 
\frac{\partial}{\partial \tilde{s}^i} (\tilde \star \tilde \Phi) = \frac43 \tilde\star \pi^1(\Phi_i) - \tilde\star \pi^{27}(\Phi_i) 
\ ,  \quad \partial_i \text{vol}(\tilde X) = \frac13 \int_7 \Phi_i \wedge \tilde \star \tilde \Phi \, ,  
\ee
where the $\pi^1$ and $\pi^{27}$ are projections to irreducible G2 representations defined  for instance in \cite{Beasley:2002db}. 
They obey the orthogonality properties 
\be
\tilde \star \pi^1(\Phi_i) \wedge \pi^{27}(\Phi_j) = 0 \ , \quad 
\tilde \star \pi^1(\Phi_i) \wedge \pi^{1}(\Phi_j) + \tilde \star \pi^{27}(\Phi_i) \wedge \pi^{27}(\Phi_j) =  \Phi_i \wedge \tilde \star \Phi_j \, . 
\ee
In practice $\pi^1(\Phi_i)$ is defined as 
\be
\label{pi1phi}
\pi^1(\Phi_i) = \left(  \frac{\int \Phi_i \wedge \tilde \star \tilde \Phi}{\int \tilde \Phi \wedge \tilde \star \tilde \Phi} \right) \tilde \Phi \, , 
\ee  
therefore we can simplify our expressions such that only $\tilde \Phi$ or $\Phi_i$ appear instead of $\pi^1(\Phi_i)$. 
Moreover, 
the $\pi^{27}(\Phi_i)$ can be also traded for $\tilde \Phi$ and $\Phi_i$ by using the manipulations 
\be
\pi^{27}(\Phi_i) \wedge B  = \Phi_i \wedge B - \pi^{1}(\Phi_i) \wedge B  
= \Phi_i \wedge B - \left(  \frac{\int \Phi_i \wedge \tilde \star \tilde \Phi}{\int \tilde \Phi \wedge \tilde \star \tilde \Phi} \right) \tilde \Phi \wedge B  \, . 
\ee 
To prove this relation one has to consider that the four-form $B$ can be expanded as $B = \tilde \star \Phi_j B^j$. 
Within this setup one can prove also that for the four-forms $A$ and $B$ we have 
\be
G^{ij}  \int \Phi_i \wedge A  \int \Phi_j \wedge B = \frac{4}{7} \int \tilde \Phi \wedge \tilde \star \tilde \Phi \int \tilde \star A \wedge B \, , 
\ee
which can be checked by expanding both $A = \tilde \star \Phi_i A^i$  and $B = \tilde \star \Phi_i B^i$. 
Notice that $G^{ij} \int \tilde \star \Phi_i \wedge \tilde \Phi  \int \tilde \star \Phi_j \wedge \tilde \Phi  =  \frac47 \left( \int \tilde \Phi \wedge \tilde \star \tilde \Phi \right)^2$. 
We can then evaluate the derivative of $P$ with respect to $\tilde s^i$ to be given by 
\be
\label{PI}
P_i  = \frac{e^y}{8} \Bigg{[}  \int \Phi_i \wedge {\cal F}_4 \, \text{vol}(\tilde X)^{- \frac37}  
-  \left(\frac{\int \tilde \star \Phi_i \wedge \tilde \Phi}{\int \tilde \star \tilde \Phi \wedge \tilde \Phi}\right) \int \tilde \Phi \wedge {\cal F}_4 \, \text{vol}(\tilde X)^{- \frac37} \Bigg{]} \, . 
\ee 
Then using the form of $P_i$ and the aforementioned properties of the G2 three-form we find  
\be 
G^{ij} P_i P_j  = 4 \left( \frac{e^y}{8} \right)^2 \int {\cal F}_4 \wedge \tilde \star {\cal F}_4 \, \text{vol}(\tilde X)^{\frac17} 
- \frac47 \left( \frac{e^y}{8} \right)^2 \left( \int \text{vol}(\tilde X)^{- \frac37}  \, \tilde \Phi \wedge {\cal F}_4 \right)^2 \, . 
\ee
We notice that the second term of this equation exactly corresponds to minus the expression in equation \eqref{Px^2}. 
Hence, from the no-scale structure \eqref{Vone} we finally find 
\be
\label{GenV}
V  =  \frac{e^{2y}}{16} \int {\cal F}_4 \wedge \tilde \star {\cal F}_4 \, \text{vol}(\tilde X)^{\frac17}\,, 
\ee
which can then be rewritten as
\be
\label{FinalV}
V = \frac{ e^{2y}}{16} \int \sqrt{\tilde g_7}  \left( e^{- x / \sqrt 7} F_4 - \gamma e^{x / \sqrt 7} \tilde \star H_3 \right)^2 \, . 
\ee
This almost concludes our proof, since, up to a factor of $1/8$ we reproduce the 10D potential (\ref{VTVR}). 
The extra 1/8 factor is due to the rescaling of the metric mentioned around equation (\ref{finalaction}).

One issue remains; the double-counting of the volume modulus, in the sense that we have been working with one scalar too much as we never enforced that the $\tilde{s}^i$ should describe fluctuations of the unit-volume G2 space.\footnote{Let us point out that a similar situation does occur in compactifications of the IIB theory where 
one has $\int \Omega \wedge \overline \Omega \sim V ||\Omega||$  as a result one may be double-counting 
the Calabi--Yau metric $g_{mn}$ volume $V$. 
This however does not happen because the volume modulus is extracted from the metric and one has $\det[g_{mn}]=1$. } 
The overall $\int \tilde \Phi \w \tilde \star \tilde \Phi$ fluctuation has to be eliminated and we would like to impose the constraint 
\be
\label{contPhi}
\int \tilde \Phi \w \tilde \star \tilde \Phi = 7 \, , 
\ee 
which equivalently means, in case of a seven-torus, that the scalars $\tilde s^i$ would be restricted to satisfy 
\be
\label{contSi}
\prod_i^7 \tilde s^i = 1 \ \quad \text{(for seven-torus)} \, . 
\ee
Both equation \eqref{contPhi} and \eqref{contSi} follow from each other in the case of a seven-torus and can be imposed on the final bosonic action to give us the correct scalar potential with the true degrees of freedom. 
However, 
at this point we have to be careful not to spoil supersymmetry, 
therefore we have to impose this constraint on the superfield level (or alternatively on the full multiplet level). 
In other words, the supersymmetry transformations have to respect the constraint \eqref{contPhi}. 
This in fact will reduce also the fermionic degrees of freedom by one. We leave the derivation of this technical point to Appendix \ref{App:constraint}.

Finally we note that, by introducing an arbitrary constant $c$, there is an infinite family of $P$-functions that gives rise to the same scalar potential $V$ 
\be
\label{SUP2}
P =  \frac{e^y}{8} \left[ \gamma e^{x/\sqrt{7}} \int \star\Phi \wedge H_3 \, \text{vol}(X)^{- \frac47} 
+ e^{-x/\sqrt{7}} \int \Phi \wedge F_4 \, \text{vol}(X)^{- \frac37}  + c \right] \, . 
\ee  
It was noticed already in \cite{Blaback:2013taa} that this is a generic way for finding ``fake superpotential'' in no-scale models. This way, any no-scale solution can be made supersymmetric in a different supergravity theory.  

\subsection{Open string moduli, axions, quantum corrections and uplifts}

We have provided a first analysis of no-scale Minkowski vacua in 3D from flux compactifications. If our ultimate goal is full moduli stabilization then we need to take care of the axions, the D2/D6 moduli (if any) and the closed string $y$-modulus. Especially the $y$ field is worrisome since it is the only one with a non-compact moduli space. In the next section we will stabilize the $y$ field using further fluxes (Romans mass). But we could equally be tempted to parallel the history of moduli stabilization in 4D as pioneered in \cite{Kachru:2003aw, Balasubramanian:2005zx, Westphal:2006tn}.  We furthermore could contemplate the further construction of de Sitter solutions from uplifting any AdS vacuum one obtains after fixing the $y$-modulus supersymmetrically (if at all possible).  In what follows we merely mention the possible paths and difficulties for achieving this.

In 4D the massless no-scale moduli are potentially fixed in a controllable fashion through quantum effects, and most notably a leading non-perturbative correction to the superpotential that involves the no-scale direction \cite{Kachru:2003aw}\footnote{Although see \cite{Sethi:2017phn} for some interesting worries about the self consistency of this approach.}. 
But this approach does not seem feasible to us in our 3D models for a simple reason: we have minimal supergravity in 3D which does not come with holomorphic protection. Hence we expect no non-renormalization theorems to exist for perturbative corrections to the superpotential, neither holomorphic arguments to restrict the form of non-perturbative corrections. This seems to rime with the fact that there seem no possible supersymmetric wrappings of Euclidean D2- or D4-branes. 
Of course quantum effects induced by the strongly coupled gauge theories on multiple D6 branes wrapping calibrated four-cycles will be there. But they are not easily computable and we think it seems realistic that we are faced with a standard Dine--Seiberg problem \cite{Dine:1985he}.

Imagine there is nonetheless a computable AdS vacuum that is sufficiently weakly coupled and with high enough masses of the moduli such that some additional ``mild'' supersymmetry breaking does not immediately destabilize the vacuum. Then we could contemplate which supersymmetry breaking sources can provide an uplift to dS, if ever. In that respect it is interesting to realize that the analogue of Klebanov--Strassler throats does exist in such backgrounds and they were constructed in \cite{Cvetic:2001ma}. Anti-D2-branes are then the natural SUSY-breaking uplift ingredient for which a probe computation \`a la \cite{Kachru:2002gs} would suggest the solutions can be metastable. However this probe computation has been refuted in \cite{Giecold:2011gw}, but that criticism in turn was argued to be essentially harmless because of the arguments in \cite{Michel:2014lva, Cohen-Maldonado:2015ssa, Armas:2018rsy, Blaback:2019ucp}, although other problems associated to anti-brane uplifting could persist as reviewed in \cite{Danielsson:2018ztv}.

\subsection{Toroidal orientifolds} 

The first step to a specific model is the choice of G2 manifold and then study the possible O2 involutions that can be defined over it. The easiest examples are G2 spaces that arise from toroidal orbifolds whose singularities are blown up or not. For singularities that can be resolved by a known geometric blow-up procedure the canonical set of examples were constructed in the original work by Joyce \cite{Joyce1, Joyce2, Joyce3}. But truly singular G2 spaces are physical as well in the context of string or M -theory and they are even required to get more interesting lower-dimensional phenomenology \cite{Acharya_2002}. Both regular and singular G2 spaces constructed from toroidal orbifolds come with extra modes not visible at the level of the torus covering space. Either these modes are really the extra moduli of cycles introduced by the geometric blow up, or they come from the twisted sector of the string. In our example below we will use the simplest singular toroidal orbifold and  be careless about the unresolved orbifold singularities, which we assume  can be resolved in string theory at the cost of extra twisted sectors. In any case the restricted set of 7 real circle radii we consider are present in most models. So in that sense we capture the ``universal'' sector of many toroidal G2 compactifications, just like the STU truncation in four-dimensional N=1 flux reductions.

The seven internal coordinates are labeled as $y^m$ 
\be
y^m \simeq y^m+1 \, . 
\ee
The finite group of isometries $\Gamma$, forming the orbifold group, should preserve the 3-form 
\be
\Phi = e^{127} - e^{347} - e^{567} + e^{136} - e^{235} + e^{145} + e^{246} \, , 
\ee 
where $e^{127} = e^1 \wedge e^2 \wedge e^7$, etc., and we have introduced the seven vielbeins of the torus 
\be
e^m = r^m dy^m \, . 
\ee
The co-associative calibration is 
\be
\star \Phi = e^{3456} - e^{1256} - e^{1234} + e^{2457} - e^{1467} + e^{2367} + e^{1357} \, . 
\ee 
For our orbifold group $\Gamma$ we use the following $\mathbb{Z}_2$ involutions 
\be
\begin{aligned}
\Theta_\alpha : (y^1, \dots, y^7 ) & \to (-y^1, -y^2, -y^3, -y^4, y^5, y^6, y^7) \, , 
\\
\Theta_\beta : (y^1, \dots, y^7 ) & \to (-y^1, -y^2, y^3, y^4, -y^5, -y^6, y^7) \, ,
\\
\Theta_\gamma : (y^1, \dots, y^7 ) & \to (-y^1, y^2, -y^3, y^4, -y^5, y^6, -y^7) \, , 
\end{aligned}
\ee
and then $\Gamma=\{\Theta_\alpha,\Theta_\beta,\Theta_\gamma\}$. 
Note that the $\Theta$ commute, they square to the identity, and they preserve the calibration $\Phi$. 
All of the three $\Theta$, the three $\Theta^2$ and the single $\Theta^3$ have each 16 copies of $\mathbb{T}^3$ as fixed points, 
but they do not act on each other freely, therefore we have a singular G2;\footnote{In \cite{Joyce2} these singularities are referred to as `bad' not because they cannot be resolved but rather because there is no straightforward prescription to do so.} 
to find the singular space of the full $\Gamma$ and to perform the blow-up is beyond the scope of our work here.

Let us focus now on the untwisted sector of this orbifold. 
This sector is somewhat universal and we wish to see how, and if, the fluxes stabilize them.   The untwisted Betti numbers 
(i.e. the Betti numbers before we resolve the singularities) are 
\be
b_0 =1 \ , \quad b_1 = 0 \ , \quad b_2=0 \ , \quad b_3=7 \, . 
\ee
These are simply found by counting the number of linearly independent p-forms $dy^{i_1 \dots i_p}$ invariant under the orbifold action $\Gamma=\{\Theta_\alpha,\Theta_\beta,\Theta_\gamma\}$. 
The seven invariant 3-forms build a basis 
\be
\label{3basis}
\Phi_i = \left( dy^{127}, - dy^{347}, - dy^{567}, dy^{136}, - dy^{235}, dy^{145}, dy^{246} \right) \ , \quad i = 1 , \dots, 7 \, , 
\ee
on which we have already expanded the calibration as $\Phi = s^i \Phi_i$, 
where the $s^i$ are the metric moduli. 
Indeed, the $s^i$ can be related to the seven torus radii $r^m$ as follows 
\be
s^1 \Phi_1 = e^{127} \ \to \ s^1 = r^1 r^2 r^7 \ , \quad  s^2 \Phi_2 = -e^{347} \ \to \ s^2 = r^3 r^4 r^7 \ , \quad  \text{etc.} 
\ee 
Then we find that 
\be 
\label{volumes}
\text{vol}(X) = \prod_{m=1}^7 r^m = \left( \prod_{i=1}^7 s^i \right)^{1/3} = \frac17 \int \Phi \wedge \star \Phi \,, 
\ee
where we use $\int_{\mathbb{T}^7} dy^1 \wedge \dots \wedge dy^7 =1$ in the covering space. 

For later convenience we also define here a basis of closed and co-closed 4-forms that are left invariant under the orbifold involutions 
\be
\label{4basis}
\Psi_i = \left( dy^{3456}, -dy^{1256}, - dy^{1234}, dy^{2457}, - dy^{1467}, dy^{2367}, dy^{1357} \right) \ , \quad i = 1 , \dots, 7 \, . 
\ee
Notice that the 3-forms \eqref{3basis} and the 4-forms \eqref{4basis} satisfy the relation (\ref{4basisGEN}) 
and that in the $\Psi_i$ basis the co-associative calibration takes the form 
\be
\label{starex}
\star \Phi = \sum_{i=1}^{7} \frac{\text{vol}(X)}{s^i} \Psi_i \, .
\ee

Let us now turn to the O-planes. As we have seen we need to include O2-planes in our setup. 
To this end consider the target space part of the O2 action, denoted $\sigma$, as the following $\mathbb{Z}_2$ involution 
\be
\sigma : (y^1, \dots, y^7 )  \to (-y^1, -y^2, -y^3, -y^4, -y^5, -y^6, -y^7) \, . 
\ee  
The $\sigma$ has $2^7$ fixed points, or alternatively different O2 sources, in the torus covering space. 
They sit at the points $y^i=0, 1/2$. 
Notice that the calibration is odd under the O2 involution 
\be
\sigma : \Phi \to - \Phi \, , 
\ee 
and that the $\Gamma$ and the $\sigma$ commute. 
We want the orbifold image of an O2 to be again some physical object, and as we will see it is an O6. 
For that it is sufficient to consider the combination of the $\Gamma$ involutions with the $\sigma$ 
\begin{align}
&\Theta_{\alpha}\sigma = (y^1, y^2, y^3, y^4, -y^5, -y^6, -y^7) \,,\\
&\Theta_{\beta} \sigma = (y^1, y^2, -y^3, -y^4, y^5, y^6, -y^7)\,,\\
&\Theta_{\gamma}\sigma = (y^1, -y^2, y^3, -y^4, y^5, -y^6, y^7)\,.
\end{align}
These 3 involutions can be interpreted as intersecting O6 planes on the positions 
\begin{align}
\begin{pmatrix} 
&{\rm O}6_{\alpha}:\quad & \times & \times & \times & \times & - & - & -  \\
&{\rm O}6_{\beta} :\quad &\times & \times & - & - & \times & \times & -  \\
&{\rm O}6_{\gamma}:\quad &\times & - & \times & - & \times & -& \times  
\end{pmatrix} \, . 
\end{align}
Here ``$\times$'' means the O6 world-volume contains these direction on the internal manifold  and ``$-$'' means the O6 positions are localized at $0,1/2$ in that direction. 
These intersections are nicely consistent with the rules for preserving supersymmetry 
and this is no coincidence because the orbifold actions were chosen such as to preserve the G2 3-form. 

One can explicitly verify the SUSY calibration of the O6 planes. According to \cite{mclean}, the condition is that a 4-cycle is calibrated if and only if $\Phi$ restricted to it vanishes identically. This is the case as can be checked for each of the O6 planes. Consider for instance the planes in the directions spanned by $e^1, e^2, e^3, e^4$. Not a single component of $\Phi$ carries indices only in that subspace. The same for the other O6 planes.

Equivalent conditions are that  $\star \Phi$ restricted to the O6 4-cycle equals exactly the volume form on it \cite{mclean}. 
Yet another condition is that the source 3-forms $j_3$ appearing in the $F_2$ Bianchi identity wedge to zero against $\Phi$. 
There is an unambiguous way to find these source forms from the involution as explained in Appendix C of \cite{Caviezel:2008ik}. When applied to our case one finds for the 3 involutions 
\be
j_{\alpha} = - e^{567}\,,\qquad  j_{\beta} = - e^{347}\,,\qquad j_{\gamma} = e^{246}\,,
\ee
which speak for themselves. We find each time that $\Phi\wedge j_3=0$.

We are still not finished with the images however. We also have 
\begin{align}
\label{Opr1}
\Theta_\alpha \Theta_\beta \sigma: y^i &\to (-y^1, -y^2, y^3, y^4, y^5, y^6, - y^7) \,,\\
\label{Opr2}
\Theta_\beta \Theta_\gamma \sigma: y^i &\to (-y^1, y^2, y^3, -y^4, -y^5, y^6, y^7)\,,\\
\label{Opr3}
\Theta_\gamma \Theta_\alpha \sigma: y^i &\to (-y^1, y^2, -y^3, y^4, y^5, -y^6, y^7) \, , \\
\label{Opr4}
\Theta_\alpha \Theta_\beta \Theta_\gamma \sigma: y^i &\to (y^1, -y^2, -y^3, y^4, - y^5, y^6, y^7)\,. 
\end{align}
In total we have 7 different directions for O6-planes 
\begin{align}
\begin{pmatrix} 
&{\rm O}6_{\alpha}:\quad & \times & \times & \times & \times & - & - & -  \\
&{\rm O}6_{\beta} :\quad &\times & \times & - & - & \times & \times & -  \\
&{\rm O}6_{\gamma}:\quad &\times & - & \times & - & \times & -& \times  \\
&{\rm O}6_{\alpha\beta}:\quad & - & - & \x & \x & \x & \x & -  \\ 
&{\rm O}6_{\beta\gamma} :\quad & - & \x & \x & - & - & \x & \x  \\ 
&{\rm O}6_{\gamma\alpha} :\quad &- & \x & - & \x & \x & - & \x  \\ 
&{\rm O}6_{\alpha\beta\gamma}:\quad &\x & - & - & \x & - & \x & \x 
\end{pmatrix} \, . 
\end{align}
These intersections are mutually supersymmetric as one can verify. This is because again all O6-planes are calibrated supersymmetrically. 
The total O6 source form that enters the Bianchi is but the sum of the forms appearing in $\Phi$ for unit value of the moduli $s^i$. 
In our current no-scale model we will cancel the O6 tadpole by simply introducing 2 D6-branes for each O6-plane. 
This adds new open string fields, i.e. gauge fields and scalar moduli. 
However, in the next section we will solve tadpoles differently and find AdS vacua instead.

In total the geometric sector contains 7 moduli that come from the seven-torus and together with the dilaton we have overall 8 real scalar moduli.  From the form of the scalar potential \eqref{GenV} we found that the vacua satisfy 
\be
\label{stabilization}
e^{- x / \sqrt 7} F_4 = \gamma e^{x / \sqrt 7} \tilde \star H_3 \, \text{vol}(\tilde X)^{- \frac17} \, . 
\ee
As before we expand the fluxes as  
\be
\label{FHfluxes}
H_3 = h^i \Phi_i  \ , \quad  F_4 = f^i \Psi_i \, , 
\ee
where $f^i$ and $h^i$ are appropriately quantized real constants restricted by the O2 tadpole
\be
\delta_{ij} \, h^i \, f^j = - Q \, . 
\ee
We want to evaluate $\tilde \star H_3$ and insert it into \eqref{stabilization}. 
This means we have to evaluate $\tilde \star \Phi_i$ in terms of $\Psi_i$. 
We find\footnote{ 
We use $\star dy^{ijk}  = \frac{\varepsilon^{ijklmnp}}{4! \sqrt{g_7}} \, g_{lq} g_{mr} g_{ns} g_{pt} \, dy^{qrst}$, 
where $\epsilon^{ijklmnp}$ is a tensor density and takes values $\varepsilon^{1234567}=1$. }
\be
\tilde \star \Phi_i = \frac{\text{vol}(\tilde X)}{(\tilde s^i)^2} \Psi_i   \,, 
\ee
where there is no summation over $i$ implied.
Now we insert everything into \eqref{stabilization} to find the seven conditions 
\be
\label{si2}
(\tilde s^i)^2  = \gamma e^{2 x / \sqrt 7} \text{vol}(\tilde X)^{6/7} \frac{h^i}{f^i}  \, \, , 
\ee
for every $i$. 
To handle \eqref{si2} we can take the product of all these seven equations together, 
i.e. evaluate $\prod_i(\tilde s^i)^2 = 1$ and use this to get a condition that their product gives $\text{vol}(\tilde X)$. 
Indeed we find, taking into account that $\gamma^7= \gamma = \gamma^{-1}$, that  
\be
\label{vdil}
x = \frac{1}{2 \sqrt 7} \text{Log}\Big{(}\gamma \prod_i \frac{f^i}{h^i} \Big{)} \, , 
\ee 
and $x$ is fixed. 
Inserting the expression for $x$ into \eqref{si2} and using the definition of the $\tilde s^i$ gives 
\be
(\tilde s^i)^2  = \Big{(}\prod_j(f^j/h^j)\Big{)}^{1/7} \, \frac{h^i}{f^i}  \, ,
\ee
for every $i$. We see that 7 out of the 8 universal moduli are fixed,  i.e. the $y$ scalar still remains undetermined and is the no-scale modulus.
 
Searching now for a supersymmetric vacuum will require that the derivatives of the superpotential with respect to the scalars $x,y,\tilde s^i$ to be zero. 
The derivatives (\ref{hPx}) and (\ref{PI}) are proportional to $\mathcal{F}_4$ and they vanish due to the Minkowski condition $\mathcal{F}_4=0$, setting no extra conditions on the fluxes. However, the derivative of the superpotential with respect to the scalar $y$ sets the following conditions
\be
P_y=P=0 \ \rightarrow \sum_i  f_i^{1/2}\Big[(\gamma h^i)^{-1/2}+(\gamma h^i)^{1/2}\Big] = 0 \, . 
\ee

\section{AdS vacua with scale separation}

\emph{``You got to keep `em separated''--- The Offspring 1994.}

\subsection{Indication for scale separation}

To find what the necessary conditions are for scale separation we apply a reasoning similar to the one in (section 4 of) \cite{Petrini:2013ika} and simply consider the dependence of the potential on the dilaton and volume modulus. This gives us necessary but non-sufficient conditions for scale separation. We first start with a discussion that is valid for compactifications down to any dimensions $D$. Consider 10D string frame and the following metric Ansatz 
\be\label{planck}
ds^2_{10} = \t_0^2 \t^{-2} ds_D^2 + \rho ds^2_{10-D} \, ,  
\ee
with $\rho^{(10-D)/2}$ the volume in 10D string frame and where 
\be
\label{dilatontr}
\t^{D-2} = \exp(-2 \phi) \rho^{\frac{10-D}{2}} \, , 
\ee
in order to find $D$-dimensional Einstein frame. In our notation $\tau_0, \rho_0$ describe the vacuum expectation values such that in a vacuum we have 
\be
ds^2_{10} =  ds_D^2 + \rho_0 ds^2_{10-D} \, . 
\ee
We use the notation in which the $(10-D)$-dimensional internal metric at unit volume is denoted $\tilde g_{10-D}$. In our paper sofar we left out the compensating $\t_0^{2}$-factor in the reduction Ansatze which effectively means we work in Planck units since
\be
S_D \supset \int_D\sqrt{g_D}\left(R_D + \ldots - V\right)\,,
\ee
where $\ldots$ represent all omissions such as kinetic terms for the scalars. 
When we use instead the formula (\ref{planck}) we find 
\be
S_D \supset \int_D\sqrt{g_D}\left(\t_0^{D-2} R_D + \ldots - \t_0^{D}V\right)\,,
\ee
such that we conclude that the Planck scale is fixed by
\be
M_p = l_p^{-1} = \t_0\,,
\ee
in string units. A proxy for the KK scale (in 10D string frame) is
$L^2_{KK} =\rho$. In a vacuum the Einstein equations tell us that
\be
R_D = \frac{D}{D-2}\, M_p^2 V\,.
\ee
So there are two length scales associated with this vacuum: 
the curvature radius $L_{\Lambda}$ and the ``vacuum energy length scale' $L_{\rho}$ defined as follows 
\be
L_{\Lambda}^{-2} = M_p^2 |V|\,,\qquad L_{\rho}^{-2}= M_p^2 |V|^{2/D} \,.
\ee
We thus have two notions of scale separation that are expressed as 
\be
\text{I} :\,\,  \frac{L_{KK}^2}{L_{\Lambda}^2} = \rho_0\tau_0^2 V\rightarrow 0\,,\qquad
\text{II}:\,\,  \frac{L_{KK}^2}{L_{\rho}^2} = \rho_0\tau_0^2 V^{2/D}\rightarrow 0\,.
\ee
The combination $\rho_0\t_0^2$ exactly equals the volume modulus $\rho^E_0$ in 10D Einstein frame. 
So if we apply the definitions to compactifications down to 3 dimensions we have simply 
\be
\text{I} :\,\,  e^{16 \beta v} V\rightarrow 0\,,\qquad
\text{II}:\,\,  e^{16 \beta v} V^{2/3}\rightarrow 0\,.
\ee

We now verify some minimal conditions for compactifications of IIA to achieve scale separation. We will use criterion I from now on. If we assume the internal space is Ricci flat (G2) and we assume O6, Romans mass, and $F_4$, $H_3$ fluxes and no net O2/D2 tension, then the scalar potential in 3D goes like 
\be \label{VVV}
V = \tfrac{1}{3!}|\tilde H|^2 \rho^{-3} \t^{-2} 
+ \tfrac{1}{4!}|\tilde F_4|^2 \rho^{-1/2} \t^{-3} 
+ |F_0|^2 \rho^{7/2} \t^{-3} 
+ T_6 \rho^{1/4} \t^{-5/2} \,.
\ee
Here the tilde symbols denote contractions with the unit-volume metric.  We hope to generate a separation of scales by cranking up the $F_4$ flux since that flux could be unbounded by tadpoles in case it wedges to zero with $H_3$. We first verify that the above ingredients are the necessary minimal requirements to find AdS vacua.  That they are sufficient will be demonstrated with an explicit example below. 

The equations of motion for stabilizing $\rho$ and $\t$; $ 
\t\,\partial_\t V = 0 = \rho\,\partial_\rho V$, can be regrouped to obtain 
\begin{align}
& 2T_6 \rho^{1/4} \t^{-5/2}  = - \tfrac{4}{3!}|\tilde H|^2 \rho^{-3} \t^{-2} - \tfrac{3}{4!}|\tilde F_4|^2 \rho^{-1/2} \t^{-3}  < 0\,,\\
& 4 |F_0|^2 \rho^{7/2} \t^{-3} = \tfrac{4}{3!}  |\tilde H|^2 \rho^{-3} \t^{-2}  + \tfrac{1}{4!} |\tilde F_4|^2 \rho^{-1/2} \t^{-3} >0 \,.
\end{align}
We learn that we need net O6 tension and non-zero Romans mass to achieve moduli stabilization with non-zero $F_4$ flux, just like in the 4D models \cite{DeWolfe:2005uu}. The on-shell potential then becomes 
\be
V = -\frac{1}{4}|\tilde F_4|^2 \rho^{-1/2} \t^{-3} <0\,, 
\ee
which is indeed AdS${}_3$. 
To verify whether scale separation could be possible we assume that we can consistently realize the following scalings which are 
compatible with the tadpole conditions 
\be
F_4 \sim N \ , \quad F_0 \sim H_3 \sim T_6 \sim N^0 \, . 
\ee 
Remarkably there is a possible scaling for the dilaton and volume with $N$ such that every term in the potential scales indeed in the same way:
\be
\t \sim N^{13/4} \ , \quad \rho \sim N^{1/2} \qquad \rightarrow\qquad V\sim N^{-8}\,,\,\, \exp(\phi) \sim N^{-3/4} \, .
\ee 
Notice that for large $N$ the modulus $\rho$ grows while the dilaton is damped so we are guaranteed to be in the supergravity limit. 
This scaling indeed implies separation since 
\be
\label{scaleVN}
\rho\tau^2 V \sim N^{-1}\,. 
\ee

\subsection{10D view on the effective theory from toroidal orbifolds}  
We now hunt for a concrete example by changing the Minkowski no-scale solution of the previous section. We add Romans mass and realize tadpoles differently and this will turn out to be sufficient.  

The following Bianchi identities lead to non-trivial RR tadpoles 
\be
\label{BIO6}
\begin{aligned}
dF_2 = & H_3 \wedge F_0 + \delta_{{\rm O}6} \, , 
\\
dF_4 = & H_3 \wedge F_2 \, , 
\\
dF_6 = & H_3 \wedge F_4 + \delta_{{\rm O}2} + \delta_{{\rm D}2}   \, , 
\end{aligned}
\ee
where for completeness we have also indicated the presence of D2-branes. 
We now take
\be
F_2 = 0 \ , \quad F_4 = F_{4A} + F_{4B} \ne 0 \ , \quad F_0 \ne 0 \ , \quad H_3 \ne 0 \, , 
\ee 
where the $F_4$ splitting refers to the way the flux wegdes with $H_3$, 
that is 
\be H_3 \wedge F_{4A} \equiv 0 \ , \quad H_3 \wedge F_{4B} = - \delta_{{\rm O}2} - \delta_{{\rm D}2} \, . 
\ee 
The $F_2$ tadpole cancellation works by cancelling the contributions from the O6-planes with $H_3 \wedge F_0$.  
We can allow for an O2 source but we assume it is canceled by a correct amount of D2-branes when $F_{4B}=0$. 
Otherwise we will not have any D2-branes and it will be $H_3 \wedge F_{4B}$ that cancels the O2 tadpole. 

The bosonic part of the 10D action that contributes to the potential is now 
\begin{equation}
\label{10Drestart}
\begin{aligned}
S =& \int_{10} \sqrt{-g_{10}} \, \left(
- \tfrac{1}{2}e^{-\phi} |H_3|^2 
-\tfrac{1}{2}e^{\phi/2} |F_4|^2  
-\tfrac{1}{2}e^{5 \phi/2} m^2 \right)  
\\
& + e^{-\phi/4}\mu' \int_{\rm O2/D2}\sqrt{-g_3}
+ e^{3\phi/4} \mu \sum_{\{\alpha,\beta,\gamma\}} \int_{{\rm O}6} \sqrt{-g_7} \, , 
\end{aligned}
\end{equation} 
where $F_0 = m$, 
and now we use the notation $\mu'=\mu_{\rm O2/D2}$ and $\mu=\mu_{\rm O6/D6}$. Our notation in the above formula for the integral over the O6 sources already anticipates our toroidal orientifold example.  In particular we use the orbifold setup that we studied in the previous section. 
Since we have calibrated O6 sources $j_{\{\alpha,\beta,\gamma\}}$, 
the $dF_2$ Bianchi (in the smeared approximation) then gives 
\be
0 =  m \, H_3 \pm \mu \, J_3 \ , \quad  J_3 = \sum \frac{j_{\{\alpha,\beta,\gamma\}}}{\text{vol}(\{\alpha,\beta,\gamma\})_3}  = \sum_i \Phi_i  \, . 
\ee 
The notation $j_{\{\alpha,\beta,\gamma\}}/\text{vol}_3$ reflects one should normalize each volume 3-form $j_3$ 
transverse to each orientifold with respect to its own 3-cycle volume. 
For example we have 
\be
\frac{j_{\alpha\beta} }{\text{vol}({\alpha\beta})_3} = \frac{e^{127}}{r^1 r^2 r^7} = \frac{s^1 \Phi_1 }{s^1} = \Phi_1 \, .  
\ee 
We take the following fluxes consistent with the tadpoles 
\begin{align}
& H_3 = h   \sum_i \Phi_i  \quad \Longrightarrow \quad h m = \pm \mu\,, \nonumber  \\ 
& F_{4A} = \sum_i f^i \Psi_i \quad \Longrightarrow \quad \sum_i f^i = 0 \,, \label{FLUXESO6}\\ 
& F_{4B} = \sum_i \hat f^i \Psi_i \quad\Longrightarrow \quad \sum_i \hat f^i = \pm \mu' / h \,. \nonumber  
\end{align}
We now compute the scalar potential for the 3D compactification and use \eqref{10DmetricV} with the internal metric given by the seven-torus orbifold. 
We split the  scalar potential $V$ into two parts, 
one that relates to the fluxes together with the O2 that we call collectively $V_\text{Flux}$ 
and one that relates to the O6-planes $V_{{\rm O}6}$. 
For the fluxes/O2 we have 
\be
\label{reflux}
V_\text{Flux} = 
\frac{1}{2}\int_7 \sqrt{\tilde{g}_7} \left[ e^{-21\beta v}\,\left(e^{-\phi/2} e^{\beta v/2} H_3\mp e^{-\beta v/2}e^{\phi/4}\tilde{\star}_7 F_4 \right)^2 
+e^{-14\beta v} e^{5 \phi/2} m^2 \right] \, , 
\ee
where $\sqrt{\tilde g_7}=1$. 
Note that all internal space contractions are with the unit-volume metric $\tilde g_{mn}$ as indicated by the tilde symbol. 
Let us now turn to the O6 contributions. For the ${\rm O}6_{\alpha\beta}$ orientifold for example we have 
\be
S_{{\rm O}6_{\alpha\beta}} 
= e^{3\phi/4} \mu \int_{{\rm O}6_{\alpha\beta}} \sqrt{-g_7} 
=  e^{3\phi/4} \mu \int_{{\rm O}6_{\alpha\beta}} \sqrt{-g_7} \  \int \frac{e^{127}}{s^1} 
= e^{3\phi/4} \mu \int \sqrt{-g_{10}} \ \frac{1}{s^1}  \, .
\ee
The contribution of all planes to the effective potential is then\footnote{We could calculate this result also by taking into account that the 
	volume form on the associated $\alpha\beta$ 4-cycle is 
	$\star j_{\alpha\beta} = \frac{\text{vol}(X)}{s^1} \Psi_1$, 
	which then would allow the following manipulations 
	$\int_{{\rm O}6} \sqrt{-g_7} = e^{3 \alpha v} \int_3 \sqrt{- \tilde g_3} \int_\text{4-cycle} \sqrt{g_4} = 
	e^{3 \alpha v} \int_3 \sqrt{- \tilde g_3} \int_\text{$\Psi^1$ 4-cycle} \star j_{\alpha\beta} = 
	e^{- 14 \beta v} \int_3 \sqrt{- \tilde g_3} (s^1)^{-1}$. 
} 
\be
V_{{\rm O}6} = - e^{3\phi/4} \mu \, e^{-14 \beta v} \, \sum_i \frac{1}{s^i} 
= - e^{3\phi/4} \mu \, e^{-17 \beta v} \sum_i \frac{1}{\tilde s^i} \, , 
\ee
where in the last step we have inserted the unit-volume fluctuations $\tilde s^i$ with the use of $s^i = e^{3\beta v} \tilde{s}^i$. 
Note that $-17 \beta v + 3\phi/4 = (-10 \beta v - \phi/2 ) + (-7 \beta v + 5 \phi/4)$ 
which are exactly the combinations of volume and dilaton that appear with $H_3$ and $m$ terms respectively, 
as one can see in \eqref{reflux}.  In the end the most convenient form for the O6 contribution is 
the one that is written in terms of the unit-volume scalars and the $x$ and $y$, 
and reads 
\be
\label{VO6unit}
V_{{\rm O}6} = 
- \mu \, e^{\frac{3}{2} y - \frac{5}{2\sqrt{7} } x} \sum_i \frac{\text{vol}(X)^{3/7}}{s^i} 
= - \mu \, e^{\frac{3}{2} y - \frac{5}{2\sqrt{7} } x} \sum_i \frac{1}{\tilde s^i}  \ , \quad \mu = \gamma \, hm \, , 
\ee
where $\gamma=\pm1$. 
Now we are ready to derive the full scalar potential $V_\text{Flux} + V_{{\rm O}6}$ from a 3D N=1 supergravity.

\subsection{The 3D supergravity}

Our aim here is to find the superpotential $P$ that defines our 3D supergravity theory.  

Taking into account that for our 3D supergravity constructions we always use the normalization $R/2$ for the Hilbert--Einstein term we have to rescale the 3D metric of the previous subsection by $1/4$. 
This means that with our superspace formulation the total scalar potential we want to reproduce is 
\be
\label{TotalO6}
\begin{aligned}
V^\text{Total} = & \frac{e^{2y}}{16} \int  \left( e^{-2 x/ \sqrt{7}} F_4 \wedge \tilde\star F_4 \, \text{vol}(\tilde X)^{\frac17} 
+  e^{2 x/ \sqrt{7}} H_3 \wedge \tilde \star H_3  \, \text{vol}(\tilde X)^{-\frac17} 
\pm 2 F_4 \wedge H_3 
\right) 
\\
& \ \ \   + \frac{ m^2}{16}  e^{ y - \sqrt{7} x  } 
- \frac{\mu}{8}  e^{\frac{3}{2} y - \frac{5}{2\sqrt{7} } x} \sum_i \frac{1}{\tilde s^i} \, , 
\end{aligned}
\ee
with the fluxes and $\mu$ given by \eqref{FLUXESO6}. 
As we have explained one should in the end 
set $\prod_i \tilde s^i=1$ to restrict to the true degrees of freedom of the toroidal orbifold. 
We will verify that our new superpotential still satisfies the conditions that allow us to impose 
the constraint on the $\tilde s^i$ without spoiling supersymmetry.

One can easily verify that in case there would only be Romans mass, the N=1 superpotential equals 
\be
\label{superR}
P^R = \frac{m}{8} \, \exp\left [ \frac12 y - \frac{\sqrt{7}}{2} x \right] \, , 
\ee
where the $R$ superscript is to stress this is the pure Romans mass contribution to the superpotential. 
Note that $P^R_i=0$. To reproduce the total scalar potential \eqref{TotalO6} we work with  the target space metric given by \eqref{scalarmetric} and we suggest that the total superpotential is simply a sum 
\be
P^\text{Total} = P + P^R \, , 
\ee
where $P$ is given is \eqref{SUP} and $P^R$ is \eqref{superR}. 
Indeed, if one writes down all the contributions to the scalar potential we have 
\be
\label{TOTOTOT}
V^\text{Total} =  G^{IJ} P_I^\text{Total} P_J^\text{Total} - 4 (P^\text{Total})^2 = V + V^R 
+ 8 P^R_x P_x + 8 P^R_y P_y - 8 P^R P \, , 
\ee
where $V=G^{IJ}P_IP_J-4P^2$, was our no-scale potential. Once we evaluate the cross-terms we find 
\be
8 P^R_x P_x + 8 P^R_y P_y - 8 P^R P \equiv V_{{\rm O}6} \, . 
\ee
Finally note that 
\be
\label{TOTOTOTcond}
\int \Phi_i \w \tilde \star \tilde \Phi \ G^{ij} \, P^\text{Total}_j = 0 \, , 
\ee
which as we explain in the appendix is the condition that 
guarantees that we can set on the superspace level $\prod_i \tilde S^i=1$ such that we reduce consistently to the true degrees of freedom.

\subsection{SUSY AdS vacua}

Now that we have found the superpotential\footnote{We now simply refer to $P^\text{Total}$ as $P$.} for our toroidal orbifold to be 
\be
\label{SPSP-TOT}
P = \frac{m}{8} e^{\frac{y}{2}  - \frac{\sqrt 7 x}{2}} 
+ \frac{\gamma h}{8} e^{y + \frac{x}{\sqrt{7}}} \sum_{i=1}^7 \frac{1}{\tilde s^i} 
+  \frac{1}{8} e^{y - \frac{x}{\sqrt{7}}} \sum_{i=1}^7  ( f^i + \hat f^i ) \tilde s^i 
\ , \quad 
\tilde s^7 = \prod_{a=1}^{6} \frac{1}{\tilde s^a} \, , 
\ee
we can look for supersymmetric vacua. We first consider what happens in the simplest case where $F_{4A} \ne 0$ and $F_{4B}=0$ 
and we take the following simple concrete set of fluxes
\be
\label{fluxesANSA}
\hat f^i = 0 \ , \quad f^i = (-f,-f,-f,-f,-f,-f,+6f) \, , \quad \gamma=1  \, , \quad f\,,h\,,m>0 \, . 
\ee
Because of the O6 tadpole the values of $h$ and $m$ are in fact very limited - all our freedom is essentially in $f$. At a later stage, once we established our solutions, we will properly quantize all fluxes and charges. With this flux choice, the superpotential simplifies to 
\be
P = - \frac{f}{8} e^{y - \frac{x}{\sqrt{7}}} \left[ \sum_{a=1}^6 \tilde s^a - 6 \prod_{a=1}^6 \frac{1}{\tilde s^a} \right] 
+ \frac{h}{8} e^{y + \frac{x}{\sqrt{7}}} \left[ \sum_{a=1}^6 \frac{1}{\tilde s^a} + \prod_{a=1}^6 \tilde s^a \right] 
+ \frac{m}{8} e^{\frac{y}{2} - \frac{\sqrt{7} x}{2}} \, . 
\ee
We search for solutions which are as isotropic as possible: meaning all $\tilde{s}^a$ ($a=1\ldots 6$) have the same value, 
which we denote $\sigma$, 
namely 
\be
\langle \tilde s^a \rangle = \sigma\,. 
\ee
The supersymmetric conditions ($\partial P =0$) become:
\be
\label{FEQS}
\begin{aligned}
0 & = a \sigma^6 + 6 \sigma + \frac{6 a}{\sigma} - \frac{6}{\sigma^6} - \frac{7b}{2} \, , 
\\
0 & = a \sigma^6 - 6 \sigma + \frac{6 a}{\sigma} + \frac{6}{\sigma^6} + \frac{b}{2} \, , 
\\
0 & = a \sigma^5 - \frac{a}{\sigma^2} - \frac{6}{\sigma^7} - 1 \, , 
\end{aligned}
\ee
where 
\be
\label{aANDb}
a = \frac{h}{f} e^{\frac{2 x}{\sqrt{7}}} \ , \quad b = \frac{m}{f} e^{- \frac{y}{2} - \frac{5x}{2 \sqrt{7}}} \, .  
\ee
One can either do a numerical integration or solve analytically 
to find 
\be
\label{numera}
a = 0.515696\dots \ , \quad b =  3.43111\dots \ , \quad \sigma = 1.32691\dots \, . 
\ee

We thus have for the dilaton and the volume modulus scalings 
\be
g_s = e^{\phi} \sim f^{-3/4} 
\ , \quad 
\text{vol}(X) = e^{7 \beta v} \sim f^{49/16} \, , 
\ee
and we can thus verify that our vacuum corresponds to weak string coupling and to large volume for large $f$. The AdS vacuum energy is given by
\be
\langle V \rangle = - \frac{1}{64 a^6 b^4} \left( 6 \sigma^2 + \frac{36}{\sigma^{12}} \right)  \frac{m^4 h^6}{f^8}   \, .  
\ee
Finally we can check the scale separation in our example
\be
|\rho \, \tau^2 \, V |  \sim f^{-1} \, , 
\ee
which matches exactly with \eqref{scaleVN}. \emph{So we find small coupling, large volume and scale separation.} We see that in our solution all of the six $\tilde s^a$ take the same numerical value, $\sigma$, by construction and the seventh is slightly different. But the torus remains, as a whole, at large values and no separate directions  get small.

Since the scalars $\tilde s^a$ are not canonically normalized the Hessian of the potential does not directly correspond to the mass matrix, but does inform us about the possible existence of tachyons (above the BF bound). 
We find 
\begin{equation}
\label{masses}
{\small
  \frac{\langle V_{IJ} \rangle }{|\langle V \rangle|}=\begin{pmatrix}
    4.282 & 3.321 & 3.321 & 3.321 & 3.321 & 3.321 & 6.823 & 2.132  \\
    3.321 & 4.282 &  3.321&  3.321& 3.321& 3.321 & 6.823 & 2.132 \\
     3.321& 3.321 & 4.282 &  3.321& 3.321& 3.321 & 6.823 & 2.132  \\
    3.321 &  3.321& 3.321 & 4.282 & 3.321& 3.321 & 6.823 & 2.132  \\
     3.321&  3.321&  3.321& 3.321 & 4.282 & 3.321 & 6.823 & 2.132\\
     3.321 &  3.321 &  3.321 &  3.321 & 3.321& 4.282 & 6.823 & 2.132 \\
     6.823&  6.823&  6.823&  6.823& 6.823& 6.823& 21.286 &  4.913 \\
    2.132 & 2.132 & 2.132 & 2.132 & 2.132& 2.132& 4.913 & 5.      \\
  \end{pmatrix} ,
  }
\end{equation}
where the lines/columns are $\tilde s^a,x,y$. 
The important thing to see in this matrix is that the 8 eigenvalues read: 39.296, 4.441, 3.434, 0.961, 0.961, 0.961, 0.961, 0.961. 
This means that all the masses are positive. 

Since we did not perform an exhaustive analysis of all possibilities, 
one can wonder if starting with different $F_{4A}$ flux values instead of \eqref{fluxesANSA} leads to more solutions. One can try for example configurations like $f^i=f (1,1,1,-1,-1,-1,0)$ or $f^i=- f (2,1,1,1,1,1,-7)$. But we were not able to find supersymmetric AdS vacua in these cases. 
We therefore postpone performing a complete analysis of the vacuum structure for a future work. 

In the appendix we explicitly show that there is an almost identical AdS$_3$ solution which can be found by taking $f\rightarrow -f$. 
This is sometimes called skew-whiffing and breaks supersymmetry \cite{Duff:1984sv}. 
For the rest the solution seems essentially the same.

\subsection{More flux}

We now briefly discuss what happens when we turn on also the $F_{4B}$ component of the $F_4$ flux. There is a good reason for doing this. In the previous solution we had to cancel the O2 tadpole with D2-branes, leaving a compact moduli space of D2 positions on the G2 space. But by adding $F_{4B}$ fluxes we can satisfy the O2 tadpole without any explicit D2 sources.

First one can think of turning on only the $F_{4B}$ component, 
in which case we will have $\hat f^i \ne 0$ and $f^i=0$. 
The $\hat f^i$ that enter the superpotential \eqref{SPSP-TOT} however are restricted by flux quantization and by the tadpole cancellation 
and thus cannot be large enough to give a scale separation. 
As a result, after one solves the eight equations $P_x=P_y=P_a=0$ 
the eight scalars will generically be stabilized in a supersymmetric AdS vacuum, 
which does not have a scale separation. 
In fact even if we turn on the $F_{4A}$ together with the $F_{4B}$ component, 
but we do not take the former to be large,  again we obtain vacua without scale separation.  We can ask how the large $F_{4A}$ component will influence the vacuum in the presence of an $F_{4B}$.  Let us assume we have some unspecified values for the $\hat f^i$ but we choose the values \eqref{fluxesANSA} for the other fluxes. 
In this case the supersymmetric conditions read
\be
\label{SUSY-genericF}
\begin{aligned}
0& = \frac{7 b}{2}  
- a \left( \sum_{a=1}^6 \frac{1}{\tilde s^a} + \prod_{a=1}^6 \tilde s^a \right) 
-  \sum_{a=1}^6 \tilde s^a 
+ 6 \prod_{a=1}^6 \frac{1}{\tilde s^a} 
+   \sum_{a=1}^6  \frac{\hat f_a}{f} \tilde s^a 
+  \frac{\hat f_7}{f} \prod_{a=1}^6 \frac{1}{\tilde s^a}   \, , 
\\
0& = \frac{b}{2}  
+ a \left( \sum_{a=1}^6 \frac{1}{\tilde s^a} + \prod_{a=1}^6 \tilde s^a \right) 
-  \sum_{a=1}^6 \tilde s^a 
+ 6 \prod_{a=1}^6 \frac{1}{\tilde s^a} 
+   \sum_{a=1}^6  \frac{\hat f_a}{f} \tilde s^a 
+  \frac{\hat f_7}{f} \prod_{a=1}^6 \frac{1}{\tilde s^a}   \, , 
\\
0& = -\frac{a}{\tilde s^b} 
+ a \prod_{a=1}^6 \tilde s^a - \tilde s^b 
- 6 \prod_{a=1}^6 \frac{1}{\tilde s^a} 
+ \frac{\hat f_b}{f} \tilde s^b 
- \frac{\hat f_7}{f} \prod_{a=1}^6 \frac{1}{\tilde s^a} \ , \quad \text{no $b$ summation} \, .  
\end{aligned}
\ee
The equations in the last line are a total of six equations, as they should be, 
because they will specify the six $\tilde s^a$. 
The $a$ and $b$ are given by \eqref{aANDb}. 
From \eqref{SUSY-genericF} it is now clear what happens in the generic case with parametrically large $F_{4A}$. 
We see that the parametrically large $f$ will damp the $\hat f^i$ contribution, 
therefore \eqref{SUSY-genericF} will essentially converge to \eqref{FEQS}, 
and we will get again a vacuum with $\tilde s^a=\sigma$ with the $a$, $b$ and $\sigma$ 
values determined by our previous solution up to negligible parametrically small corrections. 

\subsection{Flux quantization}
Finally, let us take flux quantization into account. The quantization rules, in our conventions for the 10D theory, are 
\begin{align}
& \int F_p = (2\pi)^{p-1}(\alpha')^{(p-1)/2}f_p\,,\qquad f_p\in \mathbb{Z}\,,\\
& \d F_{8-p} =\ldots + (2\pi)^7\alpha'^4Q \delta_{9-n}\,.
\end{align}
The charge of a single Dp-brane, denoted $Q_p$ is
\be
Q_p = (2\pi)^{-p}(\alpha')^{-(p+1)/2} \, , 
\ee
and the charge of an Op-plane is $-(2)^{p-5}$ that of a Dp-brane. 
So an O6-plane has the charge of 2 anti-D6-branes. 
If the charge quantum of the Romans mass is $M$ and that of the $H_3$-flux is $K$, then the integrated form of the $F_2$ Bianchi identity 
is the RR tadpole condition for each cycle 
\be \label{tadpole}
KM = 2N_{{\rm O}6} - N_{{\rm D}6}\,,
\ee
where $N_{{\rm O}6}$ and $N_{{\rm D}6}$ denote number of O6-planes and D6-branes respectively. 
There are two ways to approach this: either one works on the covering space in which a single O6 involution leaves multiple 
8 O6 fixed points in a transversal $\mathbb{T}^3$ or one works in the orbifolded space in which a single O6 has 6 other orbifold images. 
We will do the first. Then we have for each O6 involution 8 fixed points. 
These make a total of 56 O6 planes with 8 corresponding to each one of the seven cycles. 
Hence for our SUSY AdS vacuum we find (we set $\alpha'=1$) 
\begin{align}
& h =  (2\pi)^2 K\,, \quad m=(2\pi)^{-1} M\,,\quad KM = 16 \qquad f = (2\pi)^3 N\,, 
\end{align}
where $N,K,M\in\mathbb{Z}$ and we assumed no D6 branes.

\section{Outlook} 
Let us summarize what we have done in this paper. We have found the form of the 3D real superpotential $P$ for G2 compactifications of IIA supergravity with O2, O6 sources and $H_3, F_4, F_0$ fluxes to be 
\be
P =  \frac{e^y}{8} \left[e^{\tfrac{x}{\sqrt{7}}} \int \star\Phi \wedge H_3 \, \text{vol}(X)^{- \frac47} 
+ e^{-\tfrac{x}{\sqrt{7}}} \int \Phi \wedge F_4 \, \text{vol}(X)^{- \frac37}  \right] +  \frac{F_0}{8} \, e^{\tfrac12 y - \tfrac{\sqrt{7}}{2} x } \, , 
\ee
where $x$ and $y$ are specific linear combinations of the 10D dilaton and volume defined in eqs (\ref{redefs}) and $\Phi$ the G2 invariant 3-form. The 3D theory is a minimal supergravity with two real supercharges and a scalar potential given by equation (\ref{Pfunc}) 
\be
V(\phi) = G^{IJ}  P_I P_J  - 4 P^2 \, , 
\ee 
where $G_{IJ}$ is the metric on the scalar manifold. To our knowledge this is the first investigation of compactifications of type II supergravity on G2 orientifolds.\footnote{Although see \cite{Braun:2017ryx} for G2 compactifications of type II without orientifolds and fluxes. Furthermore, some of the AdS$_3$ solutions in \cite{Dibitetto:2018ftj, Dibitetto:2019yam, Legramandi:2019xqd} could be related to our findings.}

We then found two classes of (supersymmetric) solutions from the critical points of $P$ or $V$. The first class consists of no-scale Minkowski solutions and the second class of scale-separated AdS$_3$ vacua. Both were established on the same toroidal orbifold, although the RR tadpoles were solved differently each time. These types of solutions are expected to exist on general classes of G2 spaces that allow involutions for O2 and O6 planes.

The no-scale solution had one clear massless direction, the $y$-scalar, but more importantly we argued that there is no obvious obstacle in stabilizing all other moduli with non-compact moduli spaces. The scalars with compact moduli spaces are D-brane positions and Abelian vectors (axions). This is in contrast with the no-scale vacua in 4D from 3-form fluxes in IIB \cite{Giddings:2001yu} where all K\"ahler moduli are massless and to get a single massless direction one needs to restrict to special Calabi--Yau spaces with single K\"ahler moduli. 

In the AdS$_3$ solutions on the other hand we also stabilized the remaining $y$-scalar. Note that all of our vacua are at tunably small coupling and large volume. This is not any special for no-scale vacua, because of the flat $y$-direction, but it is for the AdS vacua where all non-compact scalars can be stabilized. Especially the separation of scales is an extra cherry on top of these AdS solutions.  In contrast to 4D compactifications the bridge from no-scale vacua to moduli stabilized vacua can be done in one and the same model. We simply added Romans mass and solved the tadpoles differently.

Our main focus was on demonstrating with simple examples what one expects to find in three dimensions and so for none of our examples have we delved into a detailed discussion of the twisted sector neither the axions and D-brane moduli, and this should be understood better. 
Also increasing the number of examples would be a relevant task for the future. We have summarized the classes of vacua in a schematic fashion in Fig.\,1. 
{\ }
\\[-0.3cm]
\begin{center}
	\begin{tikzpicture}[node distance=2cm]
	\label{FIG}
	\tikzstyle{arrow} = [thick,->,>=stealth]
	\tikzstyle{blob} = [rectangle, rounded corners, minimum width=3cm, minimum height=1cm,text centered, draw=black, fill=gray!02]
	\node (noscale) [blob] {No-scale};
	\node (sep1) [blob, below of=noscale, yshift=-1cm] {AdS w/ scale sep.};
	\node (nosep) [blob, right of=noscale, xshift=4.5cm] {AdS w/out scale sep.};
	\node (sep2) [blob, right of=sep1, xshift=4.5cm] {AdS w/ scale sep.};
	\draw [arrow] (noscale) -- node [anchor=east] {$F_0\ne0$}  node [anchor=west] {$F_{4A}\ne0$} (sep1); 
	\draw [arrow] (sep1) -- node[anchor=south] {$F_{4B} \ne 0$} (sep2);
	\draw [arrow] (noscale) -- node [anchor=north] {$F_{4B}\ne0$}  node [anchor=south] {$F_0\ne0$} (nosep); 
	\draw [arrow] (nosep) -- node[anchor=west] {$F_{4A} \ne 0$} (sep2);
	\end{tikzpicture}   
\end{center}
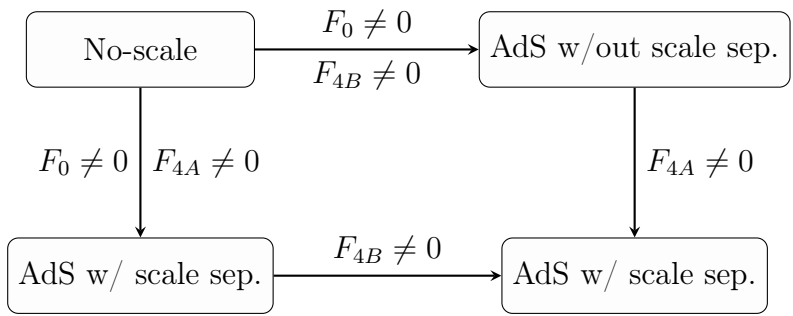
Fig.\,1:\,{\it The backgrounds/vacua depending on the form of the $F_4$ flux and the value of the Romans mass.  The parametrically large $F_{4A}$ component generates the scale separation for the AdS vacua. The $F_{4B}$ fluxes can be used to cancel O2 tadpoles.} \\

We want to emphasize that the no-scale vacua can be understood in the form of 10-dimensional solutions with localized and backreacting orientifolds \cite{Blaback:2010sj}. On the other hand, this is not understood for the AdS$_3$ vacua since they feature seven intersecting O6 planes and it is not known how to find backreacted solutions of this kind, although the recent results of \cite{Junghans:2020acz, Marchesano:2020qvg} can most likely be applied here as well and could be encouraging. 

Our main motivation for constructing the AdS${}_3$ vacua comes from holography. Already for a while there is an interest in settling the discussions about the consistency of flux vacua with scale separation. The existence of such vacua is the foundation of conventional string phenomenology. Constructing the would-be CFTs dual to scale-separated AdS vacua or show they do not exist (``bootstrap them away'') would be  the natural way forward \cite{Polchinski:2009ch, deAlwis:2014wia, Conlon:2018vov, Alday:2019qrf}. This endeavor has not yet materialized in actual concrete statements and this is why we prefer to establish a landscape of scale-separated flux vacua in 3D using the ``standard techniques'' in string phenomenology whose consistency is being debated. The reason is that 2D CFTs have been studied in more detail and especially recently some novel results seem to point against the existence of certain AdS$_3$ vacua with very high moduli stabilization (such that one arrives at pure gravity in the IR)  \cite{Benjamin:2019stq}. If the same can be argued for our scale-separated AdS$_3$ vacua it almost certainly implies 
that also 
the AdS$_4$ vacua in massive IIA \cite{DeWolfe:2005uu} neither have a holographic dual CFT \cite{Aharony:2008wz} because something is inconsistent about their construction \cite{Banks:2006hg, McOrist:2012yc}.

Finally we point out that a natural extension of our results is to include G2 torsion along the lines of \cite{DallAgata:2005zlf, Danielsson:2014ria} and check whether other scale-separated AdS or even dS solutions can arise.

\section*{Acknowledgements} 

We would like to thank Miguel Montero for discussions. 
The work of FF and TVR is supported by the KU Leuven C1 grant ZKD1118C16/16/005. This research was supported in part by the National Science Foundation under Grant No. NSF PHY-1748958.
GT thanks the ITF of KU Leuven for the hospitality during the early stages of this project.

\appendix
\section{3D minimal supergravity}\label{app:3D}
To describe the matter-coupled 3D N=1 supergravity we will follow closely notation and conventions from \cite{Andringa:2009yc}. 
The three-dimensional Clifford algebra has the $2\times2$ $\gamma$-matrices $\gamma^a$, where $a=0,1,2$ are tangent space indices, 
and the matrices satisfy $\{\gamma^a , \gamma^b \} = 2 \eta^{ab}$, 
whereas the properties of Majorana spinors in 3D can be found in \cite{Freedman:2012zz}. 
We will refer with Greek letters $\mu,\nu=0,1,2$ to world indices. 
To make our presentation easier we will emulate a superspace description in terms of two real Grassmann variables $\theta_1$ and $\theta_2$, 
even though we will not enter into a specific superspace 
construction\footnote{See for example \cite{Becker:2003wb,Kuzenko:2011xg,Buchbinder:2017qls} for a recent and full superspace presentation.}, 
instead we will follow closely the multiplet setup of \cite{Andringa:2009yc}. 
The 3D N=1 supermultiplets we will utilize are given bellow: 
\begin{itemize}
	
	\item Supergravity sector: $e_\mu^a$ is the dreibein, 
	and $\psi_\mu$ is the gravitino which is a spin-3/2 Majorana spinor. 
	This multiplet has a real scalar auxiliary field S. 
	These component fields appear in the Ricci scalar superfield 
	\be
	{\rm S} = {\rm S} + i \theta^2 \left( R + 6 {\rm S}^2 \right) + \text{fermions} \, ,  
	\ee
	where we abuse notation and use the same letter for the real superfield and for its lowest component, 
	and $R$ is the 3D Ricci scalar. 
	Note that because of the Grassmann nature of the $\theta$ we have $(i \theta^2)^*=i \theta^2$. 
	The supergravity sector appears also in the real super-density 
	\be
	{\cal E} =  e - 8 i \theta^2  e {\rm S}  + \text{fermions} \, , 
	\ee
	where $e = \sqrt{-g_3}$.

	\item Matter sector: $\phi^I$ are real scalars, 
	$\chi^I$ are spin-1/2 Majorana spinors, 
	and $F^I$ are real scalar auxiliary fields. 
	The indices $I=1, \dots n$ take values on the target space scalar manifold with real coordinates $\phi^I$ 
	and with Riemannian target space metric: $G_{IJ}(\phi)$. 
	Their superspace expansion is 
	\be
	\phi^I = \phi^I + i \theta^2 F^I + \text{fermions} \, . 
	\ee

	\item Gauge sector: The gauge fields are denoted as $A_\mu^{(A)}$ where the indices $(A)$ indicate that the field transforms in the adjoint, 
	and the gaugini are denoted by $\lambda^{(A)}$ and are Majorana spin-1/2 fermions. 
	These multiplets do not have independent auxiliary fields because the off-shell degrees of freedom of $A_\mu$ and $\lambda$ match. 
	
\end{itemize} 
From the above ingredients we can built locally supersymmetric actions, 
by using a single superspace integral $i \int d^2 \theta$ and taking into account that $i \int d^2 \theta (i \theta^2 ) =1$. 
For the supergravity sector we have 
\be
\frac{i}{2} \int d^2 \theta {\cal E} \, {\rm S} = \frac12 e R - e {\rm S}^2 + \text{fermions} \, . 
\ee
For the kinetic terms of the matter superfields we have 
\be
-\frac{i}{64} \int d^2 \theta {\cal E} \, G_{IJ}(\phi) \overline \chi^I \chi^J = 
- e G_{IJ} \partial_\mu \phi^I \partial^\mu \phi^J 
+ \frac{1}{16} G_{IJ}(\phi) F^I F^J 
+ \text{fermions} \, , 
\ee 
where $G_{IJ}(\phi)$ is the real Riemannian target space metric 
and for the superpotential $P(\phi)$ we will always use 
\be
\frac{i}{2} \int d^2 \theta {\cal E} \, P(\phi) = \frac12 e P_I F^I - 4 e P{\rm S} + \text{fermions} \, . 
\ee
Here $P(\phi)$ is a real function of the $\phi^I$ and $P_I = \partial P / \partial \phi^I$. 
Adding up these ingredients (and including the gauge sector which has no auxiliary fields) 
we can then built the most general Lagrangian for our purposes, 
which has bosonic sector 
\be
\label{sugra3DAPP}
e^{-1} {\cal L} = \frac12 R 
- g^{\mu \nu} G_{IJ}(\phi) \partial_\mu \phi^I \partial_\nu \phi^J 
- \frac14 f(\phi) F_{\mu \nu}^{(A)} F^{\mu \nu (A)} - V(\phi) \, ,  
\ee
with 
\be
V(\phi) = G^{IJ}  P_I P_J  - 4 P^2 \, , 
\ee 
and the gauge kinetic function $f(\phi)$ real but otherwise unrestricted. 
The auxiliary fields $F^I$ and S have been already integrated out in \eqref{sugra3DAPP}. 
The $G^{IJ}$ is the inverse of the target space metric $G_{IJ}$. 
For completeness let us only point out that the quadratic gravitino sector has the form 
\be
e^{-1} {\cal L}_{3/2} = - \frac12 \overline \psi_\mu \gamma^{\mu \nu \rho} D_{\nu} \psi_\rho 
- \frac12 P \, \overline \psi_\mu \gamma^{\mu \nu} \psi_\nu  \, . 
\ee
Then we verify that for SUSY-AdS$_{3}$ we have \cite{Freedman:2012zz} 
\be 
\langle P_i \rangle = 0 \ , \quad m_{3/2} = P = \frac{1}{2 L_{\rm AdS}}  \ , \quad \langle V \rangle = - 4 P^2 = - \frac{1}{L_{\rm AdS}^2}  \, . 
\ee 
The fermionic shifts on a generic maximally symmetric background are given by 
\be
\begin{aligned}
	\delta \psi_\mu \Big{|}_{\rm shift} & = \frac14 \omega_\mu^{ab} \gamma_{ab} \epsilon 
	- P \gamma_\mu \epsilon \, , 
	\\
	\delta \chi^I \Big{|}_{\rm shift} & = G^{IJ} P_J \epsilon \, , 
	\\
	\delta \lambda^{(A)} \Big{|}_{\rm shift} & = 0 \, , 
\end{aligned}
\ee
where $\epsilon$ is the 2-component fermionic Majorana local supersymmetry parameter.

\section{Enforcing the unit-volume constraint}\label{App:constraint}
The way we will reduce the independent degrees of freedom in a consistent supersymmetric  way is by using a superpspace Lagrange multiplier.  To this end we define the real 3-form superfield (not a super 3-form however)   
\be
\mathbb{\Phi} = \Phi_i \tilde S^i \, , 
\ee
where the $\tilde S^i$ are here the real superfields with lowest components $\tilde s^i$. 
We now postulate that the effective theory with the correct degrees of freedom is given by 
the Lagrangian 
\be
\label{TTT}
{\cal L} + i \int d^2 \theta \left[ {\cal E} \Lambda \left( \int_X \mathbb{\Phi} \w \tilde \star \mathbb{\Phi} \, -7 \right) \right] \, , 
\ee
where we have explicitly kept the $X$ in $\int_X \mathbb{\Phi} \w \tilde \star \mathbb{\Phi}$ to indicate that 
that integration is over the internal space, 
and $\Lambda$ is a real Lagrange multiplier superfield 
\be
\Lambda = L + \theta \chi^L + i \theta^2 F^L \, . 
\ee
Once we vary the Lagrange multiplier superfield we have 
the superspace expression 
\be
\label{super-cond}
\delta \Lambda \ : \ \int \mathbb{\Phi} \w \tilde \star \mathbb{\Phi}  = 7 \, , 
\ee
which guarantees that the condition \eqref{contPhi} is imposed consistently on the full superspace level. 
Notice that this condition indeed eliminates one scalar degree of freedom 
due to $\int \mathbb{\Phi} \w \tilde \star \mathbb{\Phi} |_{\theta=0} = \int \tilde \Phi \w \tilde \star \tilde \Phi$, 
but it will also eliminate a fermion degree of freedom because the $\theta$ order term of \eqref{super-cond} gives 
\be
\sum_j \int \Phi_j \w \tilde \star \tilde \Phi \, \chi^j = 0 \, . 
\ee
Here we have referred to the superpartners of $\tilde s^i$ as $\chi^i$.

We have until now seen that by introducing the Lagrange multiplier $\Lambda$ we can restrict to the correct number of degrees of freedom 
and keep supersymmetry. 
However, 
we want to verify that the extra term we put in the action will not alter the form of our scalar potential. 
To see this we go to the bosonic sector of \eqref{TTT} and we focus on the extra term 
\be
\label{EXTRA}
\begin{aligned}
	& i \int d^2 \theta \left[ {\cal E} \Lambda \left( \mathbb{\Phi} \w \tilde \star \mathbb{\Phi} \, -7 \right) \right] 
	\\
	& = e F^\Lambda \left( \int \tilde \Phi \w \tilde \star \tilde \Phi  -7 \right) 
	-8 e \, {\rm S} L \left( \int \tilde \Phi \w \tilde \star \tilde \Phi  -7 \right) 
	+ e L \int \Phi_j \w \tilde \star \tilde \Phi \,  F^j \, .  
\end{aligned}
\ee
Clearly when we vary $F^L$ we get \eqref{contPhi} and \eqref{contSi} as we want. 
Then notice that the second term that contains the supergravity auxiliary field S automatically drops out, 
and we are left only with the third term which crucially contains the matter auxiliary fields $F^j$.

Let us now describe how the variation of the third term in \eqref{EXTRA} works. 
First notice that this term essentially acts as a constraint once we vary $L$ that reads 
\be
\label{LFi}
\int \Phi_i \wedge \tilde \star \tilde \Phi \, F^i = 0 \, . 
\ee
This relation restricts the auxiliary fields of the $\tilde s^i$ multiplets and reduces them by one, 
consistently with the fact that we wanted to eliminate one complete multiplet. 
Since the auxiliary fields $F^i$ are now subject to a restriction, 
in order to vary them and integrate them out to derive the 
scalar potential we have to include this new constraint in the variation. 
The simplest way to do this is to study the yet un-restricted auxiliary fields sector 
\be
{\cal L}_{F^i} = \frac{1}{16} e G_{ij} F^i F^j + \frac12 e P_i F^i + e L \int \Phi_j \w \tilde \star \tilde \Phi \,  F^j \, , 
\ee
where we have kept the Lagrange multiplier $L$ and have not integrated it yet, 
such that the $F^i$ are un-restricted and we can vary them normally. 
Now we vary both the $F^i$ and $L$ to find 
\be
\label{eomFi}
F^i = - 4 \, G^{ij} P_i  - 8 L \, G^{ij} \int \Phi_j \wedge \tilde \star \tilde \Phi  \, ,  
\ee
and of course also \eqref{LFi}. 
Now we multiply \eqref{eomFi} with $\int \Phi_i \wedge \tilde \star \tilde \Phi$ to find 
\be
\label{PhiEOMFi}
\int \Phi_i \wedge \tilde \star \tilde \Phi \, F^i = - 4 G^{ij} \int \Phi_i \wedge \tilde \star \tilde \Phi \, P_i  - 224\, L  \, ,  
\ee
where we are now using the fact that vol$(\tilde X)=1$ because the $\tilde s^i$ have been already restricted from the variation of $F^L$. 
Now we observe that the left hand side term in \eqref{PhiEOMFi} is vanishing because of the constraint \eqref{LFi}, 
whereas the first term on the right hand side is also vanishing because 
an explicit calculation for our superpotentials shows that 
\be
\label{GPF}
G^{ij} P_i \int \Phi_j \wedge \tilde \star \tilde \Phi = 0 \, . 
\ee
Then from \eqref{PhiEOMFi} we see that 
\be
L = 0 \, . 
\ee 
Therefore going back to \eqref{eomFi} we see that on-shell 
\be 
F^i = -4 \, G^{ij} P_i \ \to \ {\cal L}_{F^i} = - e G^{ij} P_i P_j \, , 
\ee 
and we get the standard contribution to the scalar potential from the $\tilde S^i$ multiplets even though they are restricted. 
We conclude that our scalar potential \eqref{FinalV} is consistent and respects N=1 supersymmetry 
even when the $\tilde s^i$ are restricted to unit-volume. 
The same holds for \eqref{TOTOTOT} because of \eqref{TOTOTOTcond}.

Let us give a different and more intuitive perspective now on why we were able to reduce to unit-volume and keep the scalar potential intact. 
We will discuss explicitly the toroidal case, 
but our discussion works for the other cases as well. 
As we said we wanted to have $\prod_i \tilde S^i=1$. 
This condition could be imposed by setting $\tilde S^i = R^{3/7} T^i$ where $\prod_i T^i=1$, with $T^i$ and $R$ real scalar superfields. 
Then we could re-derive the full theory using $P=P(x,y,r,t^i)$ with this new set of superfields, 
where $R|_{\theta=0}=r$ and $T^i|_{\theta=0}=t^i$. 
The most important property of the superpotential is that we would have $\partial P / \partial r=0$, 
which would also give $F^R=0$. 
Then to reduce to unit-volume, such that we do not double count the volume, 
we would have to set on the superfield level $R=1$ which would give $r=1$ and $F^R=0$. 
The latter condition is completely compatible with the fact that $\partial P / \partial r=0$ 
and this is exactly why our scalar potential does not change form.

\section{Non-SUSY AdS$_3$}

Here we verify that there is a non-SUSY AdS$_3$ when we flip the sign of the $F_4$ flux for the SUSY solution.
We first write the scalar potential in terms of the $\tilde s^a$
\be
\label{Vxytildes}
V^\text{Total} = F(\tilde s^a) e^{2y - \frac{2x}{\sqrt{7}}} 
+ H(\tilde s^a) e^{2y + \frac{2x}{\sqrt{7}}} 
+ C\, e^{ y - \sqrt{7} x  } 
- T(\tilde s^a) e^{\frac{3y}{2}  - \frac{5x}{2\sqrt{7} } } 
\, , 
\ee
where $C = \frac{ m^2}{16} $ and 
\be
\begin{aligned}
	F(\tilde s^a) &=  \frac{f^2}{16} \left[  \sum_a (\tilde s^a)^2 
	+ 36  \prod_a \frac{1}{(\tilde s^a)^2} \right] \ , 
	\\
	H(\tilde s^a) &= \frac{h^2}{16}  \left[ \sum_a \frac{1}{(\tilde s^a)^2} + \prod_{b} (\tilde s^b)^2 \right]  \, , 
	\\ 
	T(\tilde s^a) &= \frac{hm}{8} \left[  \sum_{a} \frac{1}{\tilde s^a} + \prod_{b} \tilde s^b \right]  \, . 
\end{aligned}
\ee
Here $F(\tilde s^a)$ is related to $F_4$ flux, $H(\tilde s^a)$ is related to the $H_3$, 
the constant $C$ is related to the Romans mass which does not depend on the $\tilde s^a$, 
and the function $T(\tilde s^a)$ is related to the O-plane contribution. 
Note that all these functions are positive definite. 

To minimize the potential we again search for vacua in which $\tilde s^a = \sigma$.  We therefore calculate 
\be
\frac{\partial V}{\partial x} \Big{|}_{\tilde s^a = \sigma} = 0 
\ , \quad  
\frac{\partial V}{\partial y} \Big{|}_{\tilde s^a = \sigma} = 0 
\ , \quad  
\frac{\partial V}{\partial \tilde s^a} \Big{|}_{\tilde s^a = \sigma} = 0 \, , 
\ee
and we find after few manipulations respectively 
\be
\begin{aligned}
	0 & =  12 \left[ \sigma^2 + \frac{6}{\sigma^{12}} \right] 
	- 2 a^2 \left[ \frac{6}{\sigma^2} + \sigma^{12} \right]  
	+7 b^2 
	- 5 ab \left[ \frac{6}{\sigma} + \sigma^{6} \right] \, , 
	\\
	0 & = 12 \left[ \sigma^2 + \frac{6}{\sigma^{12}} \right]  
	+ 2 a^2 \left[ \frac{6}{\sigma^2} + \sigma^{12} \right] 
	+ b^2 
	- 3 ab \left[ \frac{6}{\sigma} + \sigma^{6} \right]  \, , 
	\\
	0 & = \left[ \sigma^2 - \frac{36}{\sigma^{12}} \right] 
	- a^2 \left[ \frac{1}{\sigma^2} - \sigma^{12} \right]  
	+ ab \left[ \frac{1}{\sigma} - \sigma^6\right] \, , 
\end{aligned}
\ee
where we see that the solution is given by the identical same values as the SUSY solution 
\be
\label{numera}
a = 0.515696\dots \ , \quad b =  3.43111\dots \ , \quad \sigma = 1.32691\dots \, . 
\ee
One can also evaluate the normalized $V_{IJ}$ on this background and see that all the eigenvalues of this matrix are positive, 
and take the values of \eqref{masses}, 
which means that all 8 scalars are stable. 
Now let us check supersymmetry. First we see that for our background 
\be
\label{P-vac}
P^T = 
\frac{3f}{4} e^{y - \frac{x}{\sqrt{7}}} \left[ \sigma - \frac{1}{\sigma^6} \right] 
+ \frac{h}{8} e^{y + \frac{x}{\sqrt{7}}} \left[ \frac{6}{\sigma} + \sigma^{6} \right] 
+ \frac{m}{8} e^{\frac{y}{2} - \frac{\sqrt{7} x}{2}} \, , 
\ee
which once we evaluate for the values \eqref{numera} we notice that 
\be
\langle V^\text{Total} \rangle \ne - 4 (P^T )^2 \, , 
\ee
which means we have a non supersymmetric vacuum.

\bibliography{refs}
\bibliographystyle{utphys}

\end{document}